\documentclass[preprint]{aastex}

\usepackage{apjfonts}

\usepackage{psfig,natbib}

\newcommand{\kms}       {km~s$^{-1}$}
\newcommand{\h}         {$h^{-1}_{70}\,$~kpc}

\newcommand{\pcm}       {cm$^{-2}$}
\newcommand{\lya}       {Ly$\alpha$}

\newcommand{\flux}      {ergs~cm$^{-2}$~s$^{-1}$~\AA$^{-1}$}

\newcommand{\cstar}     {C~II$^*$}

\lefthead{ }
\righthead{D. V. Bowen et al.}

\received{}
\accepted{}

\begin{document}

\title{A Comparison of Absorption and Emission Line Abundances
in the Nearby Damped Lyman-$\alpha$ Galaxy SBS~1543+593\altaffilmark{1}}

\altaffiltext{1}{Based on observations with the NASA/ESA
Hubble Space Telescope, obtained at the Space Telescope Science
Institute, which is operated by the Association of Universities for
Research in Astronomy, Inc., under NASA contract NAS5-26555.}

\author{David V.~Bowen$^{2}$, 
Edward B.~Jenkins$^{2}$, Max Pettini$^{3}$, Todd M.~Tripp$^{4}$} 

\affil{$\: $}

\affil{$^{2}$ Princeton University Observatory, Princeton, NJ 08544} 

\affil{$^{3}$ Institute of Astronomy, Madingley Rd., Cambridge CB3 0EZ, UK}

\affil{$^{4}$ Dept.\ of Astronomy, University of Massachusetts, Amherst,
  MA~01003}

\begin{abstract}
  
  We have used the Space Telescope Imaging Spectrograph (STIS) aboard HST to
  measure a sulfur abundance of [S/H] $=\:-0.41\pm0.06$ in the interstellar
  medium (ISM) of the nearby damped Lyman-$\alpha$ (DLA) absorbing galaxy
  SBS~1543+593. A direct comparison between this QSO absorption line abundance
  on the one hand, and abundances measured from \ion{H}{2} region emission
  line diagnostics on the other, yield the same result: the abundance of
  sulfur in the neutral ISM is in good agreement with that of oxygen measured
  in an \ion{H}{2} region 3\,kpc away.  Our result contrasts with those of
  other recent studies which have claimed order-of-magnitude differences
  between \ion{H}{1} (absorption) and \ion{H}{2} (emission) region abundances.
  We also derive a nickel abundance of [Ni/H] $\:< -0.81$, some three times
  less than that of sulfur, and suggest that the depletion is due to dust,
  although we cannot rule out an over-abundance of alpha-elements as the cause
  of the lower metallicity.  It is possible that our measure of [S/H] is
  over-estimated if some S~II arises in ionized gas; adopting a plausible star
  formation rate for the galaxy along the line of sight, and a measurement of
  the \cstar~$\lambda 1335.7$ absorption line detected from SBS~1543+593, we
  determine that the metallicity is unlikely to be smaller than we derive by
  more than 0.25 dex.  We estimate that the cooling rate of the cool neutral
  medium is $\log\:[l_c$~(ergs~s$^{-1}$~H~atom$^{-1}$)]$\:\approx\:-27.0$, the
  same value as that seen in the high redshift DLA population.

\end{abstract}


\section{Introduction}

Since the launch of the {\it Hubble Space Telescope} (HST) it has been hoped
that a comparison could be made between galactic elemental abundances deduced,
respectively, from emission line measurements made of H~II regions within some
designated 
nearby galaxy, and ultraviolet (UV) absorption line data recorded in the
spectra of a UV-bright source whose sight line passed through the same galaxy.
Both measurements would yield the composition of the present-day interstellar
medium (ISM) of such a galaxy, either in \ion{H}{2} regions ionized by OB
stars, or in neutral, diffuse, interstellar clouds which happen to lie in
front of the background source. The initial expectation has been that the two
techniques should give the same answer when applied to the same location
within a particular galaxy. Indeed, for the solar neighborhood the agreement
has now been verified empirically to a satisfactory degree of precision (see
the discussion of this point by \citealt{esteban04}).  Why, then, should we be
concerned that this may not be the case in other galaxies?

One reason arises from the puzzling result that the metallicity of damped
\lya\ systems (DLAs) --- the class of QSO absorbers with the highest column
densities of neutral gas --- is consistently below solar at essentially all
redshifts, from $z = 0.1$ to $z > 4$ \citep{kulkarni05, jx_100, pettini99}.
If DLAs were representative of the galaxy population as a whole
\citep[e.g.,][]{wolfe90},\ their element abundances should grow with time to
the near-solar levels typical of today's galaxies \citep{fukugita04}.  Radial
abundance gradients have been appealed to as a possible explanation for the
lower than solar values \citep{phl1226_radial, chen05,ellison05}.
Quantitatively, however, the magnitude of such gradients in nearby spirals has
recently been questioned \citep{bresolin04}, and remains unknown at high
redshifts.

A different aspect of the same `problem' is the possibility that the
\ion{H}{2} regions of star-forming dwarf galaxies may be self-enriched in the
nucleosynthetic products---chiefly oxygen---of the most massive, and
short-lived, stars formed during a bout of star formation.  This proposal,
originally put forward by \citet{kunth86}, has recently been reconsidered in
the light of {\it Far Ultraviolet Spectroscopic Explorer\/} (FUSE)
observations of a number of low metallicity dwarf galaxies, summarized by
\citet{cannon05}.  The FUSE data seem to show that the abundances of both
oxygen and nitrogen in the absorbing \ion{H}{1} gas are systematically lower,
by up to one order of magnitude, than those deduced from nebular emission
lines in \ion{H}{2} regions. Since nitrogen, a product of intermediate mass
stars, is released into the ISM several hundred Myr after oxygen, such an
offset may indicate the existence of an extended, lower-metallicity halo
surrounding the inner star-forming regions of dwarf galaxies, rather than
being evidence of self-pollution by individual \ion{H}{2} regions.

The seemingly straightforward consistency check of measuring abundances in the
spectrum of a QSO lying behind a galaxy with good emission-line metallicity
diagnostics has proved deceptively difficult to carry out, since the number of
suitable QSO-galaxy alignments is small: the foreground galaxy should $-$
ideally $-$ be at low redshift so that abundances in individual H~II regions
can be measured from the ground; and the background source must be bright
enough in the UV to be observed at high resolution with HST, since the
important diagnostic lines lie in that wavelength range. Although good
QSO-galaxy alignments exist, most of the UV sources lie $1-2$ magnitudes
below the limit reachable with HST to make this technique routine.

An alternative method has been to use previously identified intermediate
redshift QSO absorption systems and to measure the metallicity of the putative
absorbing galaxy. In this case, light is collected across the entire galaxy,
rather than from individual H~II regions, which are unresolved at such
redshifts.  \citet{chen05} conducted such a test in a few galaxies believed to
be the hosts of low-redshift DLAs. Again, although there are suggestions of an
offset, its origin is open to different interpretations, partly because the
element measured in absorption is iron, which exhibits a highly variable
degree of dust depletion in the interstellar medium (as well as having a
different nucleosynthetic origin from oxygen).

The metallicity of a galaxy is, of course, closely related to the formation
and evolution of its stellar population and the consequent enrichment of the
interstellar medium within the galaxy.  A recently developed line of inquiry
into the star formation process and the regulation of the interstellar gas in
DLAs has been through measurements of \cstar\ absorption lines. The ground
state of C~II is split into two fine-structure levels, and the transition from
one of these [the ($2p$) $^2 P_{3/2}$ level] to the first excited ($^2D$)
state produces a \cstar\ absorption line at 1335.7~\AA.  The population of the
fine structure levels is determined by collisions of the C~II ion with
electrons and neutral hydrogen atoms, and the line can be used to measure
important diagnostics. \citet{wolfe03a} used \cstar\ observed in $� 30$
high-$z$ DLAs to infer a star formation rate (SFR) of $�\:6\times 10^{-3}$
$M_\odot$~yr$^{-1}$~kpc$^{-2}$ for \cstar\ arising predominantly in a cold
neutral medium \citep{wolfe03b, wolfe04}. They were also able to measure a
cooling rate of $\simeq\:10^{-27.0}$ ergs~s$^{-1}$~H~atom$^{-1}$, a value one
dex smaller than that found by \citet{lehner04} for gas cooling in the Milky
Way.

In this paper we present a direct comparison between emission- and
absorption-based abundance determinations in a nearby galaxy and report that,
unlike the previous studies reviewed above, we do find excellent agreement
between the two sets of metallicity measurements.  

In a previous paper \citep{sbs1543_1} we found that the dwarf galaxy
SBS~1543+593 (at a heliocentric systemic velocity of $2868$~\kms --- see
\citealt{sbs1543_2}) produces a DLA in the spectrum of the background QSO
HS~1543+5921 ($z_{\rm em} = 0.807$; \citealt{RH98}).  The QSO sight line
intersects the inner regions of the galaxy, at only 2.4\,arcsec, or
$\simeq\:0.5\,h_{70}^{-1}$\,kpc, from its visual
center\footnote{$h\:=\:H_0/70$, where $H_0$ is the Hubble constant, and
  $q_0\:=\:-0.55$ is assumed throughout this paper. No corrections are made
  for the velocity of the Milky Way relative to the Cosmic Microwave
  Background.}. Using low-resolution spectra from HST, we found that
the galaxy produced a DLA with $N$(\ion{H}{1})\,$ = 10^{20.35}$
cm$^{-2}$.  From their analysis of the emission
line spectrum of the brightest \ion{H}{2} region located in a faint spiral arm
3.3\,kpc from the center of the galaxy, \citet{regina04} concluded that the
abundances of oxygen and nitrogen are approximately 1/3 and 1/10 solar
respectively, consistent with the low luminosity ($M_B = -16.8$) of
SBS~1543+593. In a subsequent paper, \citet{regina05} determined the sulfur
abundance of the same emission line region to be $\simeq\:0.5\pm0.4$ 
times the solar value.
They also used the absorption line data presented in this paper to compare the
metallicity of the neutral gas with that from the H~II region, and determined
that emission and absorption line measurements yield similar abundances.

In this paper, we present a detailed analysis of our data, taken with the
{\it Space Telescope Imaging Spectrograph} (STIS) aboard HST. We refine our initial
estimate of $N$(H~I), determined originally from lower resolution STIS data,
and calculate the elemental abundance of sulfur and a lower limit for the
abundance of nickel. We also investigate possible ranges for the abundances of
oxygen, nitrogen and silicon, and highlight the uncertainty in the results
which arise from using the available STIS data. Of particular concern is the
question of how much of the observed S~II absorption might actually arise in
ionized gas, and not just from the neutral component measured from the \lya\ 
line.  We perform conventional photo-ionization modeling to show that most of
the S~II comes from the same neutral component traced by the H~I; however, we
also develop a new method to calculate an ionization correction based on a
plausible measure of the star formation rate along the line of sight, and the
partial measurement of the \cstar~$\lambda 1335.7$ line found to arise in
SBS~1543+593. This method again predicts that most of the S~II arises in the
H~I absorbing medium. We conclude by comparing our abundances with those
measured in the H~II regions of SBS~1543+593, as well as those found in DLAs.
We comment briefly on the fact that the cooling rate of the cool neutral
medium in  SBS~1543+593 may well be similar to that of the DLAs.

\begin{deluxetable}{lcccl}
\tabletypesize{\small}
\tablewidth{0pt}
\tablecaption{Journal of Observations\tablenotemark{a} \label{tab_obs}}
\tablecolumns{5}
\tablehead{
\colhead{}                         &
\colhead{Exposure Time}         &
\colhead{CVZ}              &
\colhead{Central Wavelength} &
\colhead{} \\
\colhead{Date}                                &
\colhead{(s)}                               &
\colhead{Orbits}                                &
\colhead{(\AA) }                        &
\colhead{Absorption Lines Covered}                               
}
\startdata
2003 Aug 12 & 25\,250 & 5  & 1222 & \lya; \ion{N}{1}~$\lambda\lambda 1199, 1200.2, 1200.7$; Si~III~$\lambda 1206$\\
2003 Aug 10 & 39\,550 & 8  & 1272 &  \ion{S}{2}~$\lambda\lambda 1259, 1253, 1250$; Si~II~$\lambda 1260$ \\
2004 Jan 29 & 14\,400 & 3  & 1272 &  \ion{S}{2}~$\lambda\lambda 1259, 1253, 1250$; Si~II~$\lambda 1260$ \\
2004 Jan 27 & 49\,500 & 10 & 1321 & \ion{O}{1}~$\lambda 1302$; \ion{Ni}{2}~$\lambda 1317$; \ion{C}{2}~$\lambda 1334$;  \\
            &         &    &      & \ion{C}{2}${\ast}~\lambda 1335$; Si~II~$\lambda 1304$\\
\enddata
\tablenotetext{a}{All the observations were conducted under GO
  program 9784. All data were taken with the 52x0.1 aperture.}
\end{deluxetable}

\section{Data Acquisition and Analysis}

\subsection{Observations}

An ideal observation of HS~1543+5921 would use an echelle mode of STIS
to record many interstellar absorption lines simultaneously.
Unfortunately, the QSO is too faint to be observed in a reasonable
allocation of HST time.  However, HS~1543+5921 is located in the
continuous viewing zone (CVZ) for HST, which made it possible to observe using
the G140M medium-resolution grating. We used the G140M at
three wavelength settings (see Table~\ref{tab_obs}) to record 55~\AA
-wide portions of the ultraviolet spectrum using the 52x0.1 aperture.
The resolution of these observations is $\sim 20$~\kms\
\citep{stis_ihb_v7}.  To reduce the effects of fixed-pattern noise,
data were taken using the dither pattern {\tt STIS-ALONG-SLIT},
which moved the QSO along the slit in increments of
0.55 arcsec. As a result, spectra were recorded at one of three possible
positions on the detector.

In choosing which atomic transitions to cover, we were guided by the
low resolution (G140L) spectrum of the QSO
\citep{sbs1543_1} which showed that the \ion{O}{1} $\lambda 1302$ line
and the \ion{N}{1}~$\lambda\lambda 1199$, 1200.2, 1200.7 triplet
are likely to be saturated (as is usually the case in DLAs).
Furthermore, at $z_{\rm DLA} = 0.0096$, the \ion{N}{1} lines fall on
the wing of the Milky Way damped \lya\ absorption line, where the flux
is reduced to only $\sim 1\times10^{-15}$~\flux , too low to enable us
to record these features with adequate S/N.

For these reasons, our first tilt setting of the G140M grating was
chosen to record the \ion{S}{2}~$\lambda\lambda 1259$, 1253, 1250
triplet which we expected to be weak and unsaturated, yet detectable
in a reasonable exposure time.  Sulfur is a good proxy for oxygen;
like oxygen (and nitrogen) it is largely undepleted by dust in the
Galactic ISM and, being an $\alpha$-capture element, its abundance
tracks that of oxygen in Galactic metal-poor stars \citep{nissen04}.
Ionization corrections are expected to be small for both \ion{O}{1}
and \ion{S}{2} when $N$(\ion{H}{1}) is large. 

Our second tilt was set to record a relatively short exposure with the
grating centered at 1222\,\AA\ to cover the \lya\ absorption line from
SBS~1543+593, and thereby measure a precise H~I column density. The
\ion{N}{1} triplet absorption was
observed at the same time, and while the low S/N was not expected to
be a major concern for the damped \lya\ line (which covers many
resolution elements), we knew that the short exposure time would make
our observations of the N~I triplet of little use. 

Finally, we included observations centered at 1321~\AA\ to cover
the \ion{Ni}{2}~$\lambda 1317$ absorption line.  This line is of interest
because nickel is a highly depleted element in local interstellar gas.
Thus, the \ion{Ni}{2}/\ion{S}{2} ratio, when compared to its solar
value, should give an indication of the degree to which refractory
elements are depleted onto dust in the ISM of SBS~1543+593. This
grating tilt also included \ion{O}{1}~$\lambda 1302$, C~II~$\lambda
1334$, and C~II$^{*}$~$\lambda 1335$ absorption from the dwarf galaxy.
These lines were expected to be of limited use, but as we show below,
we were actually able to use them to impose some constraints on
physical conditions in the absorbing gas.

\subsection{Data extraction} 

The pipeline reduction of the spectra centered at 1222\,\AA\ proved
inadequate for the smooth removal of the geocoronal \lya\ emission
line, and provided an incorrect estimate of the
background---significant counts were observed at the bottom of the
\lya\ profiles, where the flux was expected to be zero.  Since these
problems arise from poor extraction of one-dimensional (1D) spectra
from the two-dimensional (2D) frames, we embarked on an optimal
extraction using the prescription given by \citet{horne86}. We
extracted data from the pipeline science-flatfielded (.flt) 2D files,
and used the wavelength arrays of the pipeline extracted (.x1d file)
spectra for calibration. Error arrays were also generated, using the
prescription detailed by \citet{horne86}.  Sensitivity functions were
created by dividing the counts by the flux in the pipeline-calibrated
spectra. When compared, the final optimally extracted spectra appeared
very similar to the pipeline data, except that our extraction provided:
cleaner subtraction of the geocoronal \lya\ line; the removal of the
correct amount of background; and the elimination of more hot pixels.
Final co-addition required re-sampling individual spectra to a common
wavelength scale, then adding each spectrum weighted by the inverse of
its variance.

\section{Ion Column Densities and Element Abundances}

Below, we discuss in detail all of the absorption lines of interest which are
relevant to our analysis of the metallicity of the interstellar gas in
SBS~1543+593.  In deriving ion column densities from the analysis of the
absorption lines, we adopt wavelengths and $f$-values for the transitions
from the extensive compilation by \citet{morton03}. For the Ni~II~$\lambda
1317$ oscillator strength, we use a new value determined by
\citet{ebj_ni21317}.  When element abundances are compared with their solar
counterparts, we use the recent reassessment of the solar abundance
scale by \citet{lodder03}.

\subsection{H~I \label{sect_HI} }

\begin{figure}[t]
\vspace*{-2.5cm}\centerline{\psfig
{figure=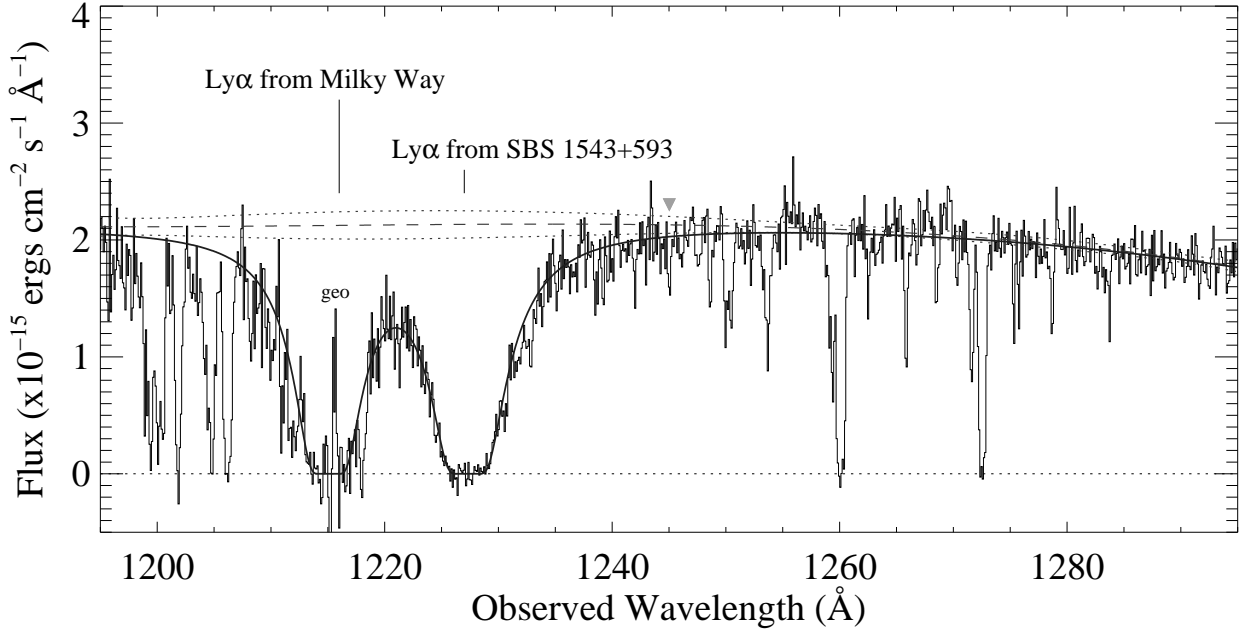,height=14cm,angle=90}}
\vspace*{-2.5cm}\caption{\small Merged STIS spectra of HS~1543+5921 taken Aug-2003, with
  the G140M grating centered at 1222~\AA\ (25.25 ksec total exposure) 
  and 1272~\AA\ (39.55 ksec;
  the gray triangle denotes the wavelength at which the data were
  joined. Data redward of this were obtained at the same epoch
  as the blue data but represent spectra from only 8 of the 11 orbits which were
  eventually used. Hence the S/N of the data redward of the triangle  
  is less than that of the final combined spectrum used for analysis of the
  absorption lines.)
  The damped H~I \lya\ lines due to absorption from both the Milky Way and
  SBS~1543+489 are indicated. Theoretical Voigt profile fits to these
  features are overplotted, along with the continuum adopted (dashed
  line) and its  $\pm 1\sigma$ deviations (dotted). For the purposes of
  this figure, the data have been rebinned two pixels into
  one. The label ``geo'' indicates residual geocoronal \lya\ emission.\label{fig_lya}}
\end{figure}

As can be seen from Figure~\ref{fig_lya}, we detect two wide damped \lya\ 
absorption lines, from the Milky Way disk at $z\simeq 0$ and from SBS~1543+393
at $z\simeq 0.009$, respectively, as expected from the lower resolution
spectrum presented in \citet{sbs1543_1}.  Unfortunately, the wavelength range
covered by the G140M grating centered at 1222~\AA\ is not wide enough to
encompass the amount of QSO continuum on either side of the two \lya\ features
which is needed to normalize the data.  We therefore used the spectra taken
with the grating centered at 1272\,\AA, which overlap the shorter wavelength
data by $\simeq\:4$~\AA, to better define the continuum.  The two sets of
observations used were obtained within a few days of each other (see
Table~\ref{tab_obs}), so we would expect no variations in the level of the QSO
flux between the two spectra.  The adopted continuum, shown in
Figure~\ref{fig_lya} along with the $1\sigma$ error to the fit, was derived
using the method described in \citet{Semb92} which uses a least-squares fit of
Legendre polynomials to intensities in regions selected either side of a
particular absorption line. This results in three normalized spectra: the
spectrum normalized by the best fit, and two other spectra normalized by the
`upper' and `lower' $1 \,\sigma$ envelopes.

To measure the \ion{H}{1} column density, $N$(\ion{H}{1}), we fitted
theoretical Voigt profiles to the data.  Details of the fitting procedures can
be found in \citet{bowen_95}.  Briefly, theoretical profiles are generated
from initial guesses of the column density, $N$, velocity, $v$, and Doppler
parameter, $b$, of the absorber. After convolution with the instrumental Line
Spread Function (LSF), the resulting line profile is compared with the
observed one; a $\chi^2$ minimization approach is then used to deduce the
best-fitting values of $N$, $b$, and $v$.  The LSF used was that appropriate
for the G140M grating at 1200\,\AA,\footnote{see
  http://www.stsci.edu/hst/stis/performance/spectral\_resolution/ } although
the \lya\ lines in the spectrum of HS~1543+5921 are so wide that the shape of
the LSF has no influence on the derivation of $N$(\ion{H}{1}).

We fitted blended profiles simultaneously to both \lya\ absorption lines from
SBS~1543+593 and from the Milky Way disk.  Wavelengths covering other narrow
absorption lines were masked out.  There are two sources of error in the
resulting values of $N$ and $v$ [The shape of the \lya\ profile is independent
of $b$ for such a large $N$(H~I)].  The first arises from errors in the
continuum fitting.  These can be estimated by refitting the spectrum as
normalized by the $1 \, \sigma$ upper and lower continua.  The second source
of error in the profile fits arises from Poisson noise. To estimate the
contribution from this source, we performed a Monte-Carlo approach, as
outlined in \cite{bowen_95}. We used the theoretical best-fit profile as a
starting point and added to this synthetic spectrum the same amount of noise
present in the data. The lines were then refitted, resulting in a new set of
values of $N$, $v$, and $b$. We repeated this process 500 times, then examined
the distributions of $N$, $b$ and $v$.  These distributions were well
approximated by Gaussians, and we used the limits which encompass 68\% of the
values as estimates of our random error in these parameters.

The result of this analysis yielded $\log N$(\ion{H}{1})$ = 20.42 \pm 0.04 \pm
0.01$ at a velocity of $2882 \pm 16$\,\kms , where the first error denotes the
uncertainty in column density arising from continuum placement, and the second
error represents that arising purely from Poisson noise. The error in $v$ from
continuum placement is negligible. These results are summarized in
Table~\ref{tab_lines}. By adding these errors in quadrature, we adopt an
\ion{H}{1} column density of $\log N$(H~I)$ = 20.42 \pm 0.04$.  The central
velocity of the \ion{H}{1} absorption can be compared to the value of
2868\,\kms\ derived from 21\,cm emission measurements \citep{sbs1543_2}.  The
difference of 14~\kms\ corresponds to approximately one detector pixel for
G140M data and is consistent with the known errors of $0.5 - 1.0$ pixels in
the absolute wavelength scale of STIS MAMA observations \citep{stis_ihb_v7}.
However, the velocity of the {\it metal} absorption lines discussed below,
2881~\kms, is very close to the \lya\ absorption velocity; since the metal
lines were recorded at different grating tilts and, in some cases, at
different epochs, the consistency in their velocities suggests that a value
close to 2881~\kms\ does indeed represent the absorption system velocity,
offset some 13~\kms\ from the systemic velocity of the galaxy, as measured
from 21~cm observations.

\subsection{\ion{S}{2}  \label{sect_SII} }

\begin{figure}
\vspace*{-1cm}\centerline{\psfig
{figure=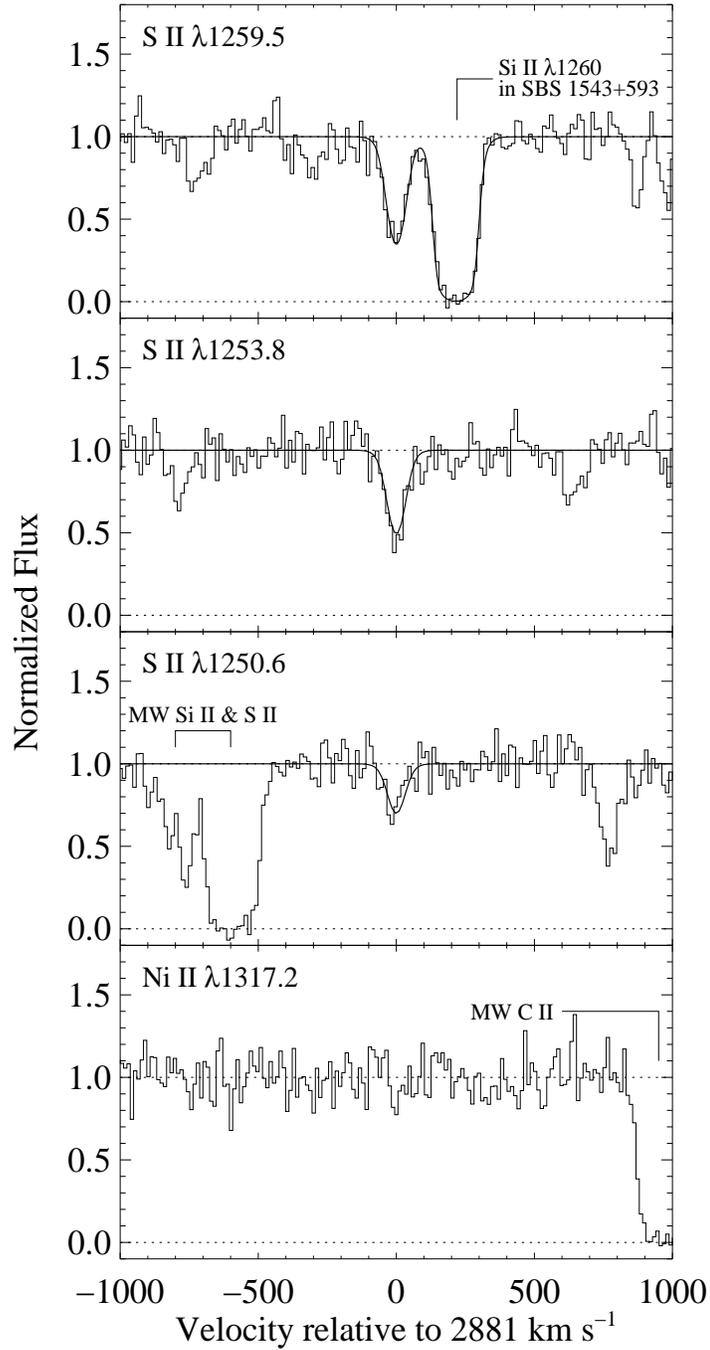,height=19cm}}
\caption{\small The top three panels show the S~II lines arising in
  SBS~1543+593. Theoretical Voigt profiles which best fit the data are
  overplotted, generated using values of $\log N$(S~II)$\:=\:15.19$,
  $b\:=\:38.5$~\kms , and $v\:=\:2881$~\kms.  
  (The S~II~$\lambda 1259$ line is blended with Si~II~$\lambda 1260$,
  which is fit with a line profile using the values given in
  Table~\ref{tab_lines}.) 
  The bottom panel shows the expected
  position of Ni~II absorption from the galaxy. Although a weak feature
  may appear to be present, it is not statistically significant. 
\label{fig_stack2}}
\end{figure}

The S~II~$\lambda\lambda 1259.52,1253.81,1250.58$ triplet is clearly
detected at the absorption redshift of SBS~1543+593 (see
Figure~\ref{fig_stack2}).  We define the velocity of the metal
absorption line
system to be $2881$~\kms\ based on the strength, S/N, and symmetry
of the Si~II~$\lambda 1260$ line (see \S\ref{sect_si2}). With $v$
fixed to this value, Voigt profile fits to the S~II triplet give
 $N$(\ion{S}{2}) $=1.55 \pm 0.14 \pm
0.08\times 10^{15}$\,\pcm\ and $b = 38.5 \pm 2.9 \pm 2.7$\,\kms . The latter
measurement suggests
that the S~II lines are resolved.
As can be seen from the breakdown in the errors, the
continuum around the \ion{S}{2} lines shows variations on scales of a
few \AA, and the resulting continuum errors are significant. Again,
by adding the errors in quadrature, we arrive at the adopted column
density of $\log N$(\ion{S}{2})~$ = 15.19 \pm 0.04$.

We can compare this value of $N$(\ion{S}{2}) with that derived using
the Apparent Optical Depth (AOD) method \citep{aod91, ebj_96}, which measures
column density as a function of velocity, $N_a$($v$).  The AOD method
provides an important check on the results from the profile fitting
analysis: if the lines of the \ion{S}{2} triplet contain any
unresolved, saturated components, the $N_a$($v$) profiles of the
stronger transitions (higher $f$-values) would appear depressed
compared to the weaker ones.  A comparison of $N_a$($v$) for each of the
three members of the \ion{S}{2} triplet shows no obvious evidence for
saturation, to within the errors for each line.  The column density
for each line, integrated over a conservative velocity range of
$-200\:<\:v\:<\:200$~\kms , is given in
the last column of Table~\ref{tab_lines}.  If we again add the
separate errors in quadrature, and weight each line by its error, we
calculate a weighted average of $\log N_a$(\ion{S}{2})~$=15.19 \pm
0.06$, which is identical to the value derived from profile fitting.

Dividing by the neutral hydrogen column density, we find
$\log N$(\ion{S}{2})/$N$(\ion{H}{1}) $ = -5.22 \pm 0.06$. 
Adopting the solar abundance 
$\log {\rm (S/H)}_{\odot} = -4.81\pm 0.04$
and taking (S/H) $=N$(\ion{S}{2})/$N$(\ion{H}{1}),
we reach the conclusion that
\[
[\rm{S/H}] = -0.41 \pm 0.06
\]
\noindent
where we have used the usual logarithmic 
notation\footnote{[X/Y] = $\log$(X/Y)  $ - \log$(X/Y)$_\odot$}. 
Hence the sulfur abundance in SBS~1543+593 is 
$0.39 \pm 0.05$ times the solar value. We discuss the validity of the 
assumption that (S/H) $=N$(\ion{S}{2})/$N$(\ion{H}{1}) in detail in
\S\ref{sect_physcondx}.

\begin{deluxetable}{lcccccccc}
\tabletypesize{\small}
\rotate
\tablewidth{0pt}
\tablecaption{Measurements of absorption lines arising from SBS~1543+593 \label{tab_lines}}
\tablecolumns{9}
\tablehead{
\colhead{}          &
\colhead{}          & 
\colhead{}          & 
\colhead{}          & \multicolumn{3}{c}{Profile fits\tablenotemark{a}} & 
\colhead{}          & 
\colhead{}         \\
\cline{5-7}
\colhead{}          & 
\colhead{$\lambda$} & 
\colhead{}          & 
\colhead{}          &   
\colhead{$v_\odot$}  &  
\colhead{$b$}       &  
\colhead{$\log N$}       &
\colhead{}          & 
\colhead{$\log N_a$\tablenotemark{a,b}}     \\
\colhead{Ion}       &   
\colhead{(\AA)}     & 
\colhead{$f$-value} &
\colhead{}          &    
\colhead{(\kms )}   & 
\colhead{(\kms)}    & 
\colhead{($\log$ [\pcm])}    &
\colhead{}          &
\colhead{($\log$ [\pcm])} 
}
\startdata
H~I      & 1215.670  & 0.4162   && 2882$\pm16$  & \nodata            & 20.42$\pm0.04\pm0.01$          && \nodata                \\
S~II     & 1259.518  & 0.0166   &&              &                    &                                && 15.18$\pm0.05\pm0.04$  \\
         & 1253.805  & 0.0109   && 2881$\pm2$   & 38.5$\pm3.0\pm2.6$ & 15.19$\pm0.04\pm0.02$          && 15.25$\pm0.10\pm0.05$  \\
         & 1250.578  & 0.0054   &&              &                    &                                && 15.18$\pm0.26\pm0.11$   \\
Ni~II    & 1317.217  & 0.0571   && \nodata      & \nodata            & \nodata                        && \phantom{1}$<\:13.83$\tablenotemark{c}  \\
O~I      & 1302.168  & 0.0486   && \nodata      & \nodata            & $\geq\:16.2$\tablenotemark{d}  && $>\:15.3$\tablenotemark{e} \\ 
Si~II    & 1260.422  & 1.1800   && \nodata      & $\leq\:41$         & $\geq\:15.1$                   && $>\:15.0 $\tablenotemark{e}\\
         & 1304.370  & 0.0863   &&              &                    &                                && \\
Si~III   & 1206.500  & 1.6300   && \nodata      & \nodata            & $\geq\:14.5$\tablenotemark{d}  && $>\:13.9$\tablenotemark{e} \\
C~I      & 1277.245  & 0.0853   && \nodata      & \nodata            & \nodata                        && \phantom{1}$<\:13.56$\tablenotemark{c} \\
C~II$^{*}$  & 1335.708 & 0.1283  && \nodata     & \nodata            & 14.28$\pm0.15$       && \nodata \\
Si~II$^{*}$ & 1264.738 & 1.0500  && \nodata     & \nodata            &      \nodata                   && \phantom{1}$\leq\:12.55$\tablenotemark{c}  \\     
\enddata
\tablenotetext{a}{Where more than one error is given, the first error indicates the uncertainty arising from continuum errors, while the
second indicates that from Poisson noise.}
\tablenotetext{b}{Column densities derived from the AOD method, with
  $N_a$($v$) integrated over $-200 < v < 200$~\kms , where $v=2881$~\kms  .}
\tablenotetext{c}{$2\sigma$ AOD column density limits, derived from continuum
  and noise errors 
  calculated over a velocity interval $-80 < v < 80$~\kms, where $v=2881$~\kms.}
\tablenotetext{d}{This lower limit comes from a fit
  to the  line assuming $b$ and $v$ equal that of the S~II line
  given in row 3.}
\tablenotetext{e}{Pixels with optical depths $\tau\:>\:3$ are set to $\tau\:=\:3$ when calculating this limit.}
\end{deluxetable}

\subsection{\ion{Ni}{2}  \label{sect_NiII} } 

The wavelength at which \ion{Ni}{2}~$\lambda 1317.217$ absorption is expected
from SBS~1543+593 (1329.85\,\AA) lies in a relatively clean part of the
spectrum. A weak feature may be present at the correct wavelength (see
Figure~\ref{fig_stack2}), but the line is not statistically significant.
We measured the AOD column density over a velocity interval of
$-80\:<v\:<\:80$~\kms\ either side of the velocity at which the Ni~II line is
expected. This velocity range corresponds approximately to the total width of the
S~II~$\lambda 1250.58$ line (see Figure~\ref{fig_stack2}). Since this
S~II transition is usually much stronger than the Ni~II~$\lambda 1317$ line,
it is highly unlikely that a Ni~II line would be wider than the observed
S~II line. 
The measured column density over this velocity interval 
is $3.23\times 10^{13}$~\pcm . In comparison, the continuum error is
$2.13\times 10^{13}$~\pcm ,  while the error from the noise is  $2.64\times
10^{13}$~\pcm . Adding these errors in quadrature yields a total $2\sigma$
error of $6.75\times 10^{13}$~\pcm , or $\log N$(Ni~II)$\:<\:13.83$. Since the
measured column density is roughly half this value, we confirm that Ni~II is
undetected at the $2\sigma$ level.

Dividing by $N$(\ion{H}{1}) as before, we find
$\log N$(\ion{Ni}{2})/$N$(\ion{H}{1}) $ < -6.59$. 
Subtracting the solar abundance 
$\log {\rm (Ni/H)}_{\odot} = -5.78 \pm 0.03$, and taking
(Ni/H)~$ =  N$(\ion{Ni}{2})/$N$(\ion{H}{1}), we obtain
\[
[\rm{Ni/H}] < -0.81 
\]

We conclude that, in the interstellar gas of 
SBS~1543+593, nickel is less abundant than sulfur by at least
a factor of 2.5.

\subsection{O~I \& N~I \label{sect_o1}}

The analysis of the \ion{O}{1}~$\lambda 1302.2$ line required some additional data
processing. The spectrum showed evidence of defective pixels or another source
of non-random noise in the core of the absorption line.
Since we shifted the target along the long axis of the slit between exposures, 
we were able to
investigate this problem by comparing individual spectra lying at different
positions on the detector. Of the eight sub-exposures of HS~1543+5921
acquired, spectra were recorded at one of three possible locations.  We found
clear evidence for two hot pixels at wavelengths close to the core of the
\ion{O}{1} line in two sub-exposures \emph{which were recorded at the same
  position on the detector} (Archive root names o8mr05030 and o8mr06030).  A
third sub-exposure (o8mr05050) also appeared to be unusually noisy (at all
wavelengths) compared to others.  Excluding these three sub-exposures from the
total improved the S/N ratio in the region of the \ion{O}{1}~$\lambda 1302$ line.

\begin{figure}[t]
\vspace*{-1cm}\centerline{\psfig
{figure=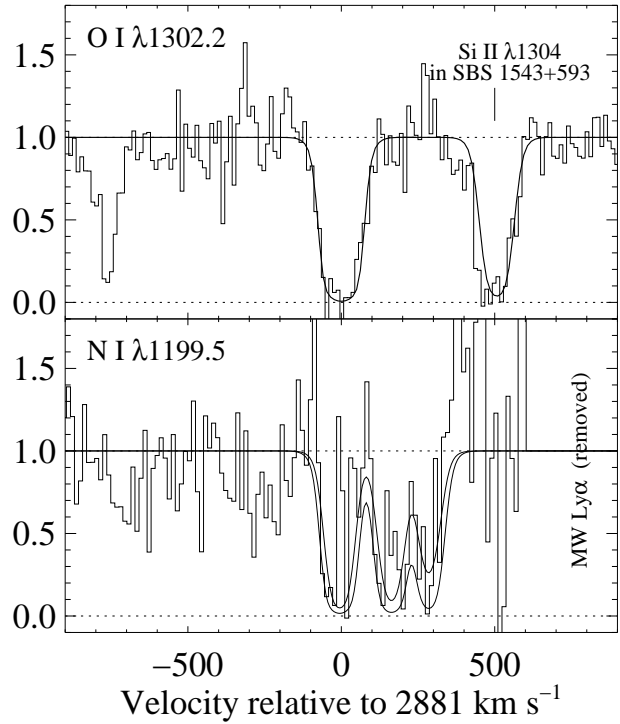,height=19cm}}
\vspace*{-8cm}\caption{\small Absorption lines arising from SBS~1543+593 for
  which, at best, only limits to ion column densities can be
  reliably derived. 
  {\bf Top:} 
  The O~I~$\lambda 1302$ line profile
  plotted over the data represents a best fit for a component with
  $b$(O~I)$\:=\:b$(S~II)$\:=\:38.5$~\kms . {\bf
    Bottom:} With the damped \lya\
  profile from the Milky Way removed, the N~I triplet
  has too low a S/N for a reliable column density to be measured. We
  overplot two sets of profiles, each with the same $b$ value as
  S~II, corresponding to the range of 
  column densities of N~I implied by the H~II region abundance 
  determinations of \citet{regina04}:
   $\log N$(N~I)~$\:=\:14.9$ and 15.4. 
\label{fig_badlines}}
\end{figure}

As expected, the \ion{O}{1}~$\lambda 1302$ line is strongly saturated
(see Figure~\ref{fig_badlines}). We used both the AOD method and Voigt
profile fitting to set lower limits to the corresponding column
density of \ion{O}{1}. In the first case, we set all pixels within the
line with optical depth $\tau \: > \:3$ to $\tau \: = \:3$, the limit beyond
which we can no longer measure the optical depth reliably.
Integrating over the same velocity interval as the \ion{S}{2}
lines, $-200\: < \: v \: < \: 200$\,\kms, we obtain $\log N_a$(\ion{O}{1})$ >
15.3$\,.

The second method involves finding the lowest column density which will
produce an acceptable fit to the observed line profile. Keeping $v$ fixed to
that of S~II, and allowing $b$ and $N$ to vary, we find that
$b\:=\:49.7$~\kms\ and $\log N$(O~I)$\:=\:15.59$.  However, there exists one
line of reasoning which suggests $N$(O~I) is larger than this.  We expect O~I
to follow H~I closely, since charge exchange locks the ionization of oxygen to
that of hydrogen \citep{field71}. On the other hand, while we expect most of
the S~II to also arise in the H~I material,
it is possible that some of the S~II might arise in H~II gas.
The velocities of these ionized components may or
may not be different from those of O~I components, but however they are
arranged in velocity, we would expect the O~I line to be as narrow, or
narrower, than the S~II line, i.e. $b$(O~I)$\:\leq\: b$(S~II).  (For this to be
true, [S/O] must be $\:\sim\:0$ in the absorbing regions, which is a reasonable
assumption considering that both species are $\alpha$-capture elements.)  This
condition is {\it not} met when $N$ and $b$ are allowed to vary for a fit to
the O~I line: as stated above, $b$(O~I)$\:=\:$49.7~\kms , whereas we measured
$b$(S~II)$\:=\:38.5$~\kms\ in \S\ref{sect_SII}.  We therefore consider that a
better limit is obtained by fixing $b$(O~I)$\:\leq\:b$(S~II)$\:=38.5\pm4.0$~\kms .
With this constraint, we find that $\log N$(\ion{O}{1})$\:\geq\:16.2$. This fit
is shown in Figure~\ref{fig_badlines}.  If $b$ is uncertain by $\pm 4$~\kms , so
that $b$(O~I)$\:=\:$ [42.5,34.5]~\kms , then $\log
N$(\ion{O}{1})$\:=\:$[15.9,16.6]. This value of $N$(\ion{O}{1}) remains a
lower limit since $b$(O~I) could in principle be narrower than $b$(S~II).

Adopting  $\log N$(\ion{O}{1})$ \:\geq\: 16.2$ as the lower limit, 
we derive $\log N$({\ion{O}{1})/$N$({\ion{H}{1})$ \:\geq\: -4.2 $. 
Since (O/H)= $N$(\ion{O}{1})/$N$(\ion{H}{1}), and 
$\log {\rm (O/H)}_\odot = -3.31\pm 0.05$, we find that the lower limit to
the oxygen abundance in SBS~1543+593 is
\[
[\rm{O/H}] > -0.9
\]

Although we believe the criterion  $b$(O~I)$\:\leq\: b$(S~II)
is a reasonable assumption, we should note that such a posit is not completely
robust. Since the O~I~$\lambda 1302$ transition is so strong, it is always
possible that weak O~I high-velocity components could be present which have no
counterparts in the S~II lines, simply because the S~II $f$-values are not
high enough to produce any absorption. These O~I components could blend with the
main profile in such as way that the measured $b$ value becomes greater than
the $b$ value of S~II. Nevertheless, as can be seen in Figure~\ref{fig_stack2}, the O~I
profile is remarkably close to the theoretical line profile for a single
component. For high velocity O~I components to widen the observed line
profile, they would have to blend symmetrically on both sides of the main
profile, and decrease in strength as their velocity increased away from the
main absorption complex.

Our data provide even fewer constraints on the N~I abundance in 
SBS~1543+593.  The bottom panel of Figure~\ref{fig_badlines} shows the
N~I triplet after removal of the Milky Way damped \lya\ profile. The
S/N is extremely low, and although absorption is present, the lines of
the triplet are ill-defined.  \citet{regina04} measured
$\log$(O/H)$\:=\:-3.8 \pm 0.2$ from the emission lines of the brightest H~II
region in the south-west outer spiral arm of the galaxy. They also
determined that $\log$(N/O) lay between $-1.7$ and $-1.2$. It follows
that $\log$ (N/H) lies between $-5.5$ and $-5.0$ for gas in the H~II
region. We can use the value of $N$(H~I) measured above, and predict
that, if the interstellar gas along the sightline had the same
metallicity as the H~II region, we should measure $\log N$(N~I)
between 14.9 and 15.4 [again assuming that (N/H) = 
$N$({\ion{N}{1})/$N$({\ion{H}{1})].
Model N~I lines, given these column densities and the same $b$ value found
for the S~II triplet, are shown plotted over the data in
Figure~\ref{fig_badlines}. We can say that the data are roughly
consistent with these column densities, but we cannot distinguish
between the upper and lower values of $\log N$(N~I).

\subsection{Si~II \& Si~III \label{sect_si2} }

A lower limit to the Si~II column density can be derived in a similar
way to that described for O~I in \S\ref{sect_o1}. Both Si~II~$\lambda
1260.4$ and 1304.4 lines are detected in the ISM of SBS~1543+593:
the 1304.4~\AA\ line has the smaller $f$-value and is the least
saturated of the two, so should give a more reliable column density;
the 1260.4~\AA\ line is more saturated, but its inclusion in the
process of fitting a theoretical line profile ensures that the $b$
value does not become too wide. With $b$ and $N$ allowed to
vary, we find $\log N$(Si~II)$\:\geq\:15.1$ and
$b$(Si~II)$\:\leq\:41$~\kms .

We also note the detection of Si~III in absorption from
SBS~1543+593 which appears in the red wing of the damped \lya\
absorption profile from the Milky Way. The
line is recorded at low S/N, but is of some use in providing information on the
ionization state of the gas, discussed in \S\ref{ion_corr}
below. Using the AOD method to set a conservative lower limit, we arrive at
$\log N_a$(Si~III)$\: > \: 13.9$, after again setting all points with optical
depth $\tau
\: > \:3$ to be $\tau\:=\:3$. A Voigt profile fit to the line keeping $v$
fixed at 2881~\kms\ and letting $b$ and $N$ vary produces the same Si~III
column density. 
We can also set
$b$(Si~III)$\:\leq\:b$(S~II) as we did for O~I; we then find $\log
N$(Si~III)$\: \geq \: 14.5$. Although the data are of too low a quality to
provide a rigorous estimate of the Si~III column density, the similarity of
these results suggests that our derived lower limit to $N$(Si~III) may be robust.

\subsection{C~I \label{sect_c1}}

As we discuss in \S\ref{sect_physical}, even though no C~I~$\lambda 1277$ is
detected from SBS~1543+593, a limit on the C~I column density is of interest
for deriving limits on the electron density within the H~I absorbing gas. The
$2\sigma$ column density limit, again measured over the velocity interval of
$-80\:<\:v\:<\:80$~\kms, and combining noise and continuum errors in
quadrature, is $3.64\times10^{13}$~\pcm , or $\log N$(C~I)$\:<\:13.56$; the
measured column density in the same interval is only $1.58\times10^{13}$~\pcm
.

\begin{figure}[t]
\vspace*{-1cm}\centerline{\psfig
{figure=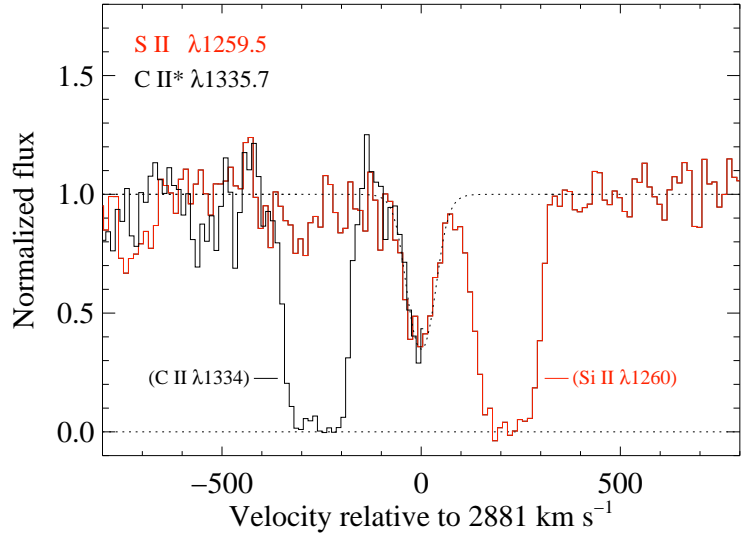,height=8cm}}
\vspace*{0cm}\caption{\small The C~II$^{*}\:\lambda 1335.7$ 
absorption line (black,solid)
  overplotted on the S~II~$\lambda 1259.5$ line (red,solid), both of which arise in
  the ISM of SBS~1543+593. Although the C~II$^{*}$ line falls right at the
  edge of our spectrum, the portion which remains suggests
  that the line is of similar strength and shape as the S~II
  line. (Other interstellar lines from the ISM of SBS1543+593 are
  labelled.) 
  A theoretical \cstar\ 
line profile with the same $b$ and $v$ derived for the S~II triplet,
  generated with $\log N$(C~II$^{*}$)$\:=\:14.28$ (see
  Table~\ref{tab_lines}) is shown as a black dotted line. \label{fig_c2star}}
\end{figure}

\subsection{C~II$^*$ \label{sect_c2star}}

The C~II$^*$~$\lambda 1335$ line arising from SBS~1543+593 lies at the very
edge of the spectrum taken with the G140M grating centered at 1321~\AA.
Unfortunately, only half of the line is recorded (see
Figure~\ref{fig_c2star}). Nevertheless, C~II$^*$ is extremely valuable in
providing insights on the conditions inside the gas clouds which give rise to
the S~II and C~II absorption lines.

Figure~\ref{fig_c2star} shows that the portion of the C~II$^*$
absorption we observed has a strength and shape that are virtually
identical to the corresponding part of the S~II~$\lambda 1259.5$ line.
If the velocity structures of the two species are the same, then so are the 
optical depths of each line, and we can state that
$N({\rm C~II}^*)=(f\lambda)_{1259.5}N({\rm
  S~II})/(f\lambda)_{1335.7}=10^{14.28}$, even if the two lines are
somewhat saturated\footnote{For the C~II$^*$ absorption, we assume
  that the 1335.663 and 1335.708$\,$\AA\ transitions are equivalent to
  a single transition with $\log f\lambda=2.234$.}. A Voigt profile
with this column density, and $v$ and $b$ fixed to that of the S~II
lines, fits the visible line profile quite accurately, as can be seen
in Fig~\ref{fig_c2star}. We can estimate an approximate error on
$N$(C~II$^*$) by seeing how the fit changes as $N$ changes, again
keeping $v$ and $b$ fixed to the S~II values. Although this is
somewhat more uncertain than a fit to an entire line profile, we
estimate that $N$(C~II$^*$) is accurate to within $\pm 0.15$ dex.

\section{Physical conditions in the ISM of SBS~1543+593 \label{sect_physcondx}}

\subsection{Ionization corrections to the neutral gas: CLOUDY modeling \label{ion_corr}}

In the previous sections, we assumed that ionization corrections
can be neglected in the derivation of abundances. To test whether this
is a reasonable assumption, we have assembled photoionization models
as described in \citet{tmt03} using CLOUDY (v96; see
\citealt{ferland98} for a general description of the code) with the
\ion{H}{1} fixed at log $N$(\ion{H}{1}) = 20.42 and [M/H] = $-0.41$.
Since the sight line obviously passes through the disk of SBS~1543+593
and starlight is readily apparent nearby (see Figure~2 in
\citealt{sbs1543_1} or Figure~1 in \citealt{regina04}), we have
assumed that the gas is photoionized mainly by starlight in the disk
of a galaxy. To approximate the ionizing flux, we have assumed that
the shape of the radiation field is similar to that in the diffuse ISM
in the disk of the Milky Way, and we have employed the Galactic disk
radiation field recently presented by \citet[][see the ``0 kpc'' curve
shown in their Figure~8]{fox05}. Given the large number of H~II
regions distributed across the disk of SBS~1543+593, this seems like a
suitable model.  It is difficult to estimate the intensity for
normalization of the ionizing flux field, but since the predicted
columns are primarily dependent on the ionization parameter $U$
($\equiv$ ionizing photon density/total H number density $=
n_{\gamma}/n_{\rm H}$), the main effect of changing the normalizing
flux intensity is to change the corresponding $n_{\rm H}$ implied by
the model; the relative column densities remain the same when
$n_{\gamma}$ and $n_{\rm H}$ are scaled up or down by the same amount.

Figure~\ref{fig_cloudyplot} compares the photoionization model column
densities to observational constraints including the measured
$N$(\ion{S}{2}) and the $2\sigma$ upper limit on $N$(\ion{Ni}{2}).  We
also show the \ion{Si}{3} column
obtained by fitting the \ion{Si}{3} profile with the $b-$value allowed
to freely vary (lower value) and with $b$ forced to have the same
value as the \ion{S}{2} lines (upper value in
Figure~\ref{fig_cloudyplot}).  Even the higher value for
$N$(\ion{Si}{3}) could significantly underestimate the true column due
to saturation and an uncertain $b-$value.

\begin{figure}[t]
\vspace*{-2cm}\centerline{\psfig
{figure=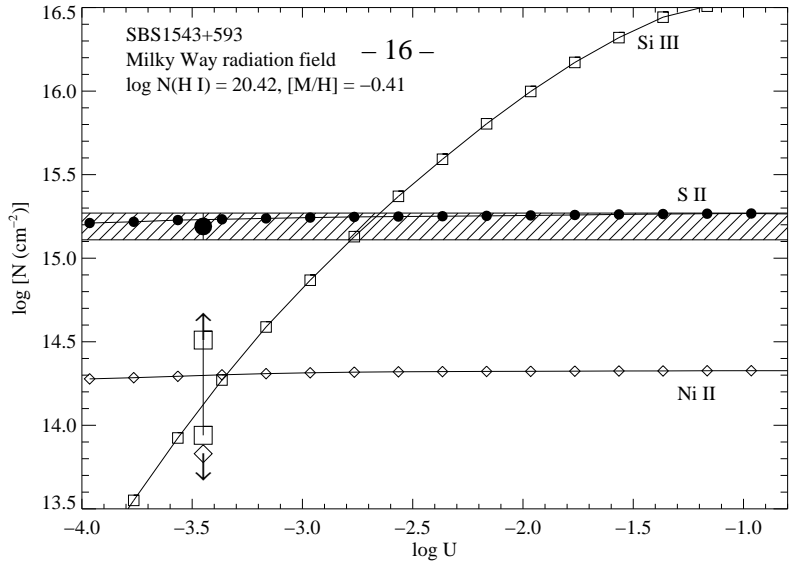,height=8cm}}
\vspace*{0cm}\caption{\small Column densities of \ion{S}{2} 
(filled circles), \ion{Ni}{2}
(open diamonds), and \ion{Si}{3} (open squares) predicted by a CLOUDY
photoionization model for a gas slab with log $N$(\ion{H}{1}) = 20.42,
overall metallicity [M/H] = $-0.41$, and solar relative abundances of
S, Ni, and Si. The ionizing radiation field was assumed to be similar
to the radiation field estimated by Fox et al. (2005) for the diffuse
ISM in the disk of the Milky Way (see the ``0 kpc'' curve shown in
Figure 8 of Fox et al.).  The model column densities are shown with
small symbols and are plotted as a function of the ionization
parameter $U$. The observed column densities are shown with larger
symbols at log $U \approx -3.45$; the hatched region indicates the
range of log $N$(\ion{S}{2}) consistent with the observed \ion{S}{2}
column within $\pm 2\sigma$. Two values for $N$(\ion{Si}{3}) are shown
corresponding to different assumptions about the Si~III
$b-$value; conservatively, these should both be treated as lower
limits (see text).\label{fig_cloudyplot}}
\end{figure}

Figure~\ref{fig_cloudyplot} shows two interesting results.  First,
even though there can be substantial amounts of ionized gas in a gas
slab with the SBS~1543+593 \ion{H}{1} column that is photoionized by
starlight (as revealed by the \ion{Si}{3} curve), the ionization
corrections for \ion{S}{2} and \ion{Ni}{2} are likely to remain rather
small or negligible.  The reason for this is that while the ion
fractions $f$ (averaged over the thickness of the gas slab) of
\ion{H}{1}, \ion{S}{2}, and \ion{Ni}{2} steadily decrease as $U$
increases, over the entire range of $U$ shown in
Figure~\ref{fig_cloudyplot}, $f$(\ion{H}{1}) $\approx \ f$(\ion{S}{2})
$\approx f$(\ion{Ni}{2}), so that ionization corrections are not
necessary (see eqn. 2 in Tripp et al. 2003).  Indeed, the strong
\ion{Si}{3} absorption detected in the SBS~1543+593 DLA is entirely
consistent with a simple photoionized gas slab, i.e., there is no need
based on the current data to invoke a more complex multiphase model.
The \ion{Si}{3} arises in the ionized
surface layer of the gas slab, but most of the \ion{H}{1}, \ion{S}{2},
and \ion{Ni}{2} is located in the shielded interior of the
slab. Second, the underabundance of \ion{Ni}{2} is probably not an
ionization effect.  Over a wide range of $U$, we find little change in
$N$(\ion{S}{2})/$N$(\ion{Ni}{2}), and at all values of $U$ that we
have considered, the model predicts a significantly higher \ion{Ni}{2}
column than the $2\sigma$ upper limit.

We must note, however, that this model is not unique. It is
straightforward to construct a UV radiation field which is much softer than
that described above; for example, one in which the flux field is
dominated by photons from an ensemble of cooler, B-type stars. In this
model, hydrogen is ionized over a longer path length than for the
harder flux field, and some S~II arises in H~II gas. Whether this UV
field is applicable for SBS~1543+593 is unclear, however; since
over 30 candidate H~II regions are identified by \citet{regina04} in
the the galaxy, it seems likely that some contribution to the
background must come from hot O-type stars. Nevertheless, it is a concern that
the sightline to the background QSO passes near the center of the galaxy,
somewhat distant from the most intense areas of star formation.

A final caveat centers on the fact that it may simply be
inappropriate to use a model which assumes that a gas slab is
illuminated by a background radiation field, which is the calculation that
CLOUDY performs. In particular, the state of the ISM illuminated from
stars and H~II regions embedded in a disk may be different from that
calculated by CLOUDY. 

Given these uncertainties, in the following section we develop an alternative
method for estimating how much S~II may actually arise in ionized gas.

\subsection{An Alternate Method for Determining the Ionization
Correction}\label{sect_alternate}

Our ultimate objective is to determine the fraction
\begin{equation}\label{F_HII}
F _{\rm H~II}({\rm S~II})\:=\:{N({\rm S~II})_{\rm H~II}\over N({\rm
S~II})_{\rm H~I}+N({\rm S~II})_{\rm H~II}}
\end{equation}
so that we can apply a multiplicative correction $\Big[1-F_{\rm
H~II}({\rm S~II})\Big]$ to our measured value of $N$(S~II) and thereby
determine the correct abundance of sulfur [S/H] in SBS~1543+593.
One way to do this is to compare
two different methods for deriving the expected emission measure (EM) of
the ionized gas $\int n(e)^2 dl$ in the H~II regions that are created by
radiation from young stars during active star formation.  The
first procedure, discussed in \S\ref{sect_SFR_HI}, starts with an
estimate for the star formation rate (SFR) per unit area, $\Sigma_{\rm
SFR}$, based on our measured $N$(H~I)
combined with a widely applicable empirical
relationship between $\Sigma_{\rm SFR}$ and gas surface density.  On the
assumption that little if any of the ionizing radiation from the stars
is lost, we can can convert $\Sigma_{\rm SFR}$ into an expectation for
the EM (\S\ref{sect_EM}).

The other, independent method for deriving the EM is through our
observation of \cstar (\S\ref{sect_c2star}).  However, the
formulation we develop for the EM, derived in \S\ref{sect_cstarHII}, involves a
number of previously unknown quantities,
such as the amount of multiply-ionized C
in the H~II regions, the apportionment of \cstar\ between H~I and H~II
regions, and, finally, the quantity we wish to determine, $\Big[1-F_{\rm
H~II}({\rm S~II})\Big]$.  By studying the production of \cstar\ in H~I
regions of our own Galaxy, we conclude in \S\ref{sect_FHII_cstar} that
only a small fraction of the \cstar\ in SBS~1543+593 should come from H~I
regions, after we compensate for differences in the star formation rates
and dust grain contents in the two systems.  Making use of this
information, we equate the two different determinations of the EM in
\S\ref{Y_cstar(C)} and find that we can set a useful upper limit for the
amount of S~II that arises from H~II regions.

\subsubsection{Star Formation Rate Based on $N$(H~I)}\label{sect_SFR_HI}

Unfortunately, it is difficult for us to determine the SFR near the line
of sight by traditional means (e.g., stellar brightness and spectral
energy distribution, the strength of H$\alpha$ emission, or the far
infrared brightness) because we are blinded by the flux from the quasar. 
Instead, we make use of a recent refinement \citep{kennicutt98b} of the
Schmidt law \citep{schmidt59} that expresses an empirical relationship
between the SFR per unit area $\Sigma_{\rm SFR}$ and the surface density
of (atomic plus molecular) hydrogen $\Sigma_{\rm gas}$
\begin{equation}\label{Schmidt_Law}
\Sigma_{\rm SFR}\:=\:2.5\pm 0.7\times 10^{-4}\:\left({\Sigma_{\rm
gas}\over 1{\rm M}_\odot\:{\rm pc}^{-2}}\right)^{1.4\pm0.15}\:{\rm
M}_\odot\:\:{\rm yr}^{-1}\:\:{\rm kpc}^{-2}
\end{equation} 
that seems to apply to a broad range of galaxy types and locations
within galaxies (including nuclear starbursts associated with molecular
gas disks).  We see no reason to believe that there could be a breakdown
of this relationship caused by either prominent bars or activity related
to galaxy mergers or tidal effects.  Drawing from experience in our own
Galaxy \citep{savage77} and the Magellanic Clouds \citep{tumlinson02},
we expect that for our observed value of $N$(H~I) the contribution from
molecular hydrogen is unlikely to exceed a few percent of the total
hydrogen.  Thus, for $N({\rm H~I})\:=\:10^{20.42\pm 0.04}\,{\rm
cm}^{-2}$, we obtain
\begin{equation}\label{Sigma_SFR}
\Sigma_{\rm SFR}\:=\:7^{+9}_{-4}\times 10^{-4}\:{\rm M}_\odot\:\: {\rm
yr}^{-1} \:\:{\rm kpc}^{-2}\:,
\end{equation}
where the stated limits include the effects from the two errors shown in
Eq.~\ref{Schmidt_Law} (which are assumed to be uncorrelated), a natural
rms dispersion of $\pm 0.3\,$dex of real outcomes on either side of the
Schmidt law, and our uncertainty in $N$(H~I), all combined in
quadrature.

\subsubsection{Anticipated Emission Measure}\label{sect_EM}

Assuming that there is no escape of Lyman Limit photons and that they
are not absorbed by dust, we can calculate the value of the EM from
$\Sigma_{\rm SFR}$.  For a \citet{salpeter55} IMF and solar abundances,
the hydrogen-ionizing photon production rate $Q({\rm H}^0)$ in
units of ${\rm s}^{-1}$ is given by
\begin{equation}\label{SFR_Q}
Q({\rm H}^0)\:=\:{\rm SFR}/1.08\times 10^{-53}\:{\rm M}_\odot\: {\rm
yr}^{-1}\:{\rm s}
\end{equation}
\citep{kennicutt98a}.  Certain factors could change the constant in
Eq.~\ref{SFR_Q}.  For instance, an IMF defined by \citet{scalo86} would
increase the constant by about a factor of three 
\citep{kennicutt98a}.  We acknowledge that the stars in SBS~1543+593 
will have compositions below solar, but the change in
ionizing fluxes should be very small \citep{lanz03}.

If all of the ionizing photons are consumed by the
ionization of H, $Q({\rm H}^0)$ per unit area must balance the product
of EM and the hydrogen recombination coefficient $\alpha^{(2)}$ to all
levels higher than the ground electronic state.  For a temperature in
the vicinity of $10^4\,$K (or $T_4=T/10^4\,{\rm K}\approx 1$) a
power-law fit to the values of $\alpha^{(2)}$ and temperature given by
\citet{spitzer_bookp107} yields
\begin{equation}\label{alpha(2)}
\alpha^{(2)}=2.55\times 10^{-13}\,T_4^{-0.833}\:\:\:{\rm cm}^3\:{\rm
s}^{-1}\, .
\end{equation}
Thus, we find that
\begin{equation}\label{EM1}
{\rm EM}\:=\:1.23\times 10^4\,T_{4,n(e)^2}^{0.833}\:\Sigma_{\rm
SFR}\:\:{\rm cm}^{-6}\:{\rm pc}
\end{equation}
where, as in Eq.~\ref{Sigma_SFR}, $\Sigma_{\rm SFR}$ is expressed in the
units ${\rm M}_\odot\,{\rm yr}^{-1}\,{\rm kpc}^{-2}$, and for an
inhomogeneous H~II region $T_{4,n(e)^2}$ is the temperature (in units of
$10^4\,$K) weighted by the local squares of the electron densities along
the line of sight, i.e., $T_{4,n(e)^2}=\int T_4n(e)^2dl\Big/\int
n(e)^2dl$.

\subsubsection{${\rm C~II}^*$ Contributions from H~II
Regions}\label{sect_cstarHII}

Collisions with electrons excite singly ionized carbon atoms to the
upper fine-structure level of the ground electronic state, creating
\cstar.  The balance between this excitation and radiative decay leads
to an equilibrium equation from which one can derive the electron
density
\begin{equation}\label{ne1}
n(e)\: =\: {g_1 A_{2,1} T^{0.5}\exp(91{\rm K}/T)\over
8.63\times 10^{-6} \Omega_{1,2}}\left( {n({\rm C~II}^*)\over n({\rm
C~II})}\right)\:=\:18.3\, T_4^{0.5}\exp(0.0091/T_4)\:\:\left( {n({\rm
C~II}^*)\over n({\rm C~II})}\right) \:\:\:{\rm cm}^{-3}\, ,
\end{equation}
where the lower level's statistical weight $g_1=2$, the upper level's
spontaneous decay rate $A_{2,1}=2.29\times 10^{-6}\,{\rm s}^{-1}$ 
\citep{nussbaumer81}, and the collision strength
$\Omega_{1,2}=2.90$ at $T=10^4\,$K \citep{hayes84}.   This
equation arises from a simplified form of the equilibrium equation in
the limit where collisional de-excitations can be neglected, which
happens when $N({\rm C~II}^*)/N({\rm C~II})\ll 1$ (which is true here)
-- see, e.g., \citealt{ebj00}.  Henceforth, we will drop the
$\exp(0.0091/T_4)$ term since it is so close to 1.0 when $T_4\sim 1$,
which seems justified for H~II regions in systems with metal contents
somewhat below that of our Galaxy, such as those in M101
\citep{kennicutt03}.  In most circumstances,
optical pumping of the excited level is unimportant  
\citep{spitzer75, sarazin79}.

An alternate expression for $n(e)$ within the H~II region is given by
the equation
\begin{equation}\label{ne2}
n(e)=\left({{\rm H}\over {\rm C}}\right)n({\rm C~II})\:y({\rm C~II})
\end{equation}
where
\begin{equation}\label{yc}
y({\rm C~II})\:=\: {n({\rm C~II})+n({\rm C~III})+n({\rm C~IV})\over
n({\rm C~II})}
\end{equation}
In an H~II region where the ionization by starlight dominates over that
from an intergalactic field, the expected values of $y({\rm C~II})$ are
strongly dependent on the average temperature of the
exciting stars $T_*$: models of ion abundances in H~II regions evaluated
by \citet{stasinska90}
indicate that the ratio of \\
$\int n({\rm C_{total}})\: n(e)\:dV/\int
n({\rm C~II})\:n(e)\:dV$
varies from 1.26 for $T_*=32,500$\,K, to 8.5
for $T_*=40,000$\,K, and could reach as high as 22 for $T_*=50,000$\,K.

It is reasonable to assume that [S/C]~$\approx$~0 in SBS~1543+593, which
permits us to  arrive at an estimate for the H to C abundance ratio,
\begin{equation}\label{H/C}
\left({{\rm H}\over {\rm C}}\right)\:=\:\left({{\rm S}\over {\rm
C}}\right)_\odot ~ {N({\rm H~I})\over \Big[1-F_{\rm H~II}({\rm
S~II})\Big] N({\rm S~II})}
\end{equation}
which we consider to be a universal abundance ratio everywhere.  We can
now formulate a useful expression for $n(e)^2$ by taking the product of
the expressions in Eqs.~\ref{ne1} and \ref{ne2},
\begin{equation}\label{EM2a}
{\rm EM}\:=\: 5.93\times 10^{-18}\,\left({{\rm H}\over {\rm
C}}\right)\,\int n({\rm C~II}^*)\,T_4^{0.5}y({\rm C~II})\,dl \:\:{\rm
cm}^{-6}{\rm pc}
\end{equation}
To be rigorous, we must allow for the possibility that $T_4$ and $y({\rm
C~II})$ could vary inside the H~II region(s), and therefore we define a
product
\begin{equation}\label{YT}
Y_{n({\rm C~II}^*)}({\rm C~II})\: T_{4,n({\rm C~II}^*)}^{0.5}\:=\:\int
T_4^{0.5}\:y({\rm C~II})\:n({\rm C~II}^*)\:dl\Bigg/ \int n({\rm
C~II}^*)\:dl
\end{equation}
so that we can work with a more streamlined representation
\begin{eqnarray}\label{EM2b}
{\rm EM}&\:=\:&5.93\times 10^{-18}\left({{\rm S}\over {\rm
C}}\right)_\odot ~ \left({N({\rm H~I})\over N({\rm S~II})}\right) ~
\times \nonumber\\
&&\left({Y_{n({\rm C~II}^*)}({\rm C~II})\: T_{4,n({\rm
C~II}^*)}^{0.5}F_{\rm H~II}({\rm C~II}^*)\over 1-F_{\rm H~II}({\rm
S~II})}\right) ~ N({\rm C~II}^*)\:\:{\rm cm}^{-6}{\rm pc}
\end{eqnarray}
after we make the substitution for (H/C) given in Eq.~\ref{H/C}. 
[$F_{\rm H~II}({\rm C~II}^*)$ is defined in a manner identical to that
for S~II in Eq.~\ref{F_HII}.]

Were it not for the uncertainties in the quantities $Y_{n({\rm
C~II}^*)}({\rm C~II})$, $T_{4,n({\rm C~II}^*)}^{0.5}$, and $F_{\rm
H~II}({\rm C~II}^*)$, we could determine $F_{\rm H~II}({\rm S~II})$ by
equating the EM given in Eq.~\ref{EM2b} with that given earlier in
Eq.~\ref{EM1}.  To provide useful constraints on these unknown
quantities, we must evaluate the probable contribution of \cstar\ from
H~I regions.

\subsubsection{An Evaluation of $F_{\rm H~II}({\rm
C~II}^*)$}\label{sect_FHII_cstar}

The discussion presented in the previous subsection focused on the
properties of fully ionized gas regions, as revealed by C~II$^*$, but
with the retention of an unknown fractional quantity $F_{\rm H~II}({\rm
C~II}^*)$ that expressed
how much of the observed C~II$^*$ actually came from such regions.  We
now examine the plausibility that $F_{\rm H~II}({\rm C~II}^*)$ could be
appreciably lower than
1.0, and that a reasonable fraction $\Big[1-F_{\rm H~II}({\rm
C~II}^*)\Big]$ of the C~II$^*$ arises from
H~I regions, as opposed to H~II regions.  One way to do this in
principle is to measure $N$(Si~II$^*$)/$N$(C~II$^*$) \citep{howk05},
since the excitation energies of the two are very different. 
Unfortunately, our 2$\sigma$ column density limit 
derived at the
expected
position of the Si~II$^*\:\lambda 1264$ line 
is $\log N$(Si~II$^*$)$\:<\:12.55$.  This limit is
not sufficiently sensitive to discriminate between contributions
from  H~II regions with
$T_4\sim 1$ and H~I regions at much lower temperatures.

We turn now to a different, but less direct way to measure $F_{\rm
H~II}({\rm C~II}^*)$.  For the
H~I regions that intersect our line of sight, the average cooling rate
per H atom arising from the emission of radiation at $157.7\, \mu$m
caused by the radiative decay of C~II$^*$ corresponds to 
\begin{eqnarray}\label{l_C}
l_C\:&=\:&(A_{2,1}hc/\lambda)\Big[1-F_{\rm H~II}({\rm
C~II}^*)\Big]N({\rm C~II}^*)/N({\rm H~I})\nonumber \\
&=\:&2.88\times
10^{-20}\Big[1-F_{\rm H~II}({\rm C~II}^*)\Big]N({\rm C~II}^*)/N({\rm
H~I})\:\:\: {\rm erg~s}^{-1}\:{\rm
H~atom}^{-1}\, .
\end{eqnarray}
For our measurements of $N$(C~II$^*$) and $N$(H~I), we obtain $\log\Big(
l_C/\Big[1-F_{\rm H~II}({\rm C~II}^*)\Big]\Big)\:=\:-25.68$.  If we omit the
factor $\Big[1-F_{\rm H~II}({\rm C~II}^*)\Big]$, this value is about the
same as the rate $\log l_C=-25.70$ (+0.19, -0.35 for the $1\sigma$
dispersion of
individual determinations) found by \citet{lehner04} for low-velocity
clouds and some accompanying ionized material in the local region of our
Galaxy.

For densities $n({\rm H})\gtrsim 0.1\,{\rm cm}^{-3}$, within the ISM of
our Galaxy the heating rate from energetic photoelectrons liberated from
dust grains dominates over those arising from cosmic ray and X-ray
ionizations \citep{wolfire95, weingartner01b}.
To a first order of approximation, we expect the dust grain heating to
scale in proportion to both (1) the local starlight density, with
special emphasis on the far-UV photons, and (2) the density of dust
grains per H atom.  This relationship is valid as long as there is no
appreciable loss of efficiency caused by the grains acquiring a positive
charge because the recombination rate with free electrons is too low --
see Fig.~16 of 
\citet{weingartner01b}.  A reasonably good gauge of
the local far-UV starlight flux is the SFR.  We can propose that the
grain density should scale in proportion to the overall metal abundance,
$10^{\rm[M/H]}$, although we caution that there are indications from IR
emission that the character of dust in low metallicity systems, and in
particular the polycyclic aromatic hydrocarbons (PAHs) which are
important for heating the gas, is different from that of our Galaxy
\citep{draine05, engelbrecht05}.  Thus, there may not be a simple downward
scaling of the abundances of all the different kinds of dust.  Nevertheless, if
we overlook this possible complication
and use our local region of the Milky Way (MW) as a
comparison standard, we can express a proportionality
\begin{equation}\label{proportionality}
{l_C({\rm SBS~1543+593})\over \Sigma_{\rm SFR}{\rm
(SBS~1543+593)}}=10^{[{\rm M/H}]}{l_C({\rm
MW})\over \Sigma_{\rm SFR}{\rm (MW)}}\, ,
\end{equation}
where $l_C({\rm SBS~1543+593})$ is defined in Eq.~\ref{l_C},
$\Sigma_{\rm SFR}{\rm (SBS~1543+593)}$
is calculated in Eqs.~\ref{EM1} and \ref{EM2b}, and $10^{[{\rm
M/H}]}=\Big[1-F_{\rm H~II}({\rm S~II})\Big]10^{-0.41}$ (see
\S\ref{sect_EM}).  For $l_C({\rm MW})$ we adopt a representative value
of $1.3\times 10^{-26}{\rm erg~s}^{-1}\:{\rm H~atom}^{-1}$, which
\citet{lehner04} found for low velocity clouds that
have sufficient column densities to make the relative contributions of
C~II$^*$ from ionized regions relatively small (see their Fig.~8).  An
acceptable range for $\Sigma_{\rm SFR}$(MW) in our region of the Galaxy
is $(3.5-5)\times
10^{-3}\, {\rm M}_\odot\, {\rm yr}^{-1}\,{\rm kpc}^{-2}$ \citep{rana91}. 
Using these
quantities, we solve Eq.~\ref{proportionality} to find that
\begin{equation}\label{F_HII_cstar_a}
F_{\rm H~II}({\rm
C~II}^*)\:=\:\Big[1+0.057T_{4,n(e)^2}^{-0.833}\,T_{4,n({\rm
C~II}^*)}^{0.5}\,Y_{n({\rm C~II}^*)}({\rm C~II})\:\Big]^{-1}
\end{equation}
With some loss of rigor for nonuniform regions where both $T_4$ and
$y({\rm C~II})$ might vary, we can consolidate the two temperature terms
with the different weightings into a single representation $T_4$ to
obtain
\begin{equation}\label{F_HII_cstar_b}
F_{\rm H~II}({\rm C~II}^*)\:=\:\Big[1+0.057T_4^{-0.333}Y_{n({\rm
C~II}^*)}({\rm C~II})\:\Big]^{-1}
\end{equation}

\subsubsection{An Evaluation of $Y_{n({\rm C~II}^*)}({\rm
C~II})$}\label{Y_cstar(C)}

As we stated earlier (\S\ref{sect_cstarHII}), the value of the inverse
of the fraction of C atoms in the singly ionized form, $Y_{n({\rm
C~II}^*)}({\rm C~II})$, can be slightly greater than 1 or very large,
depending on the hardness of the ionizing radiation (i.e., the
characteristic temperature of the exciting stars).  Any constraints that
we can place on the value of this parameter are critical for our being
able to estimate how much of the \cstar\ arises from H~II regions (see
Eq.~\ref{F_HII_cstar_b} above), and this quantity is essential for our
final estimate of $F_{\rm H~II}({\rm S~II})$.

If we convert the $\Sigma_{\rm SFR}$ derived from the Schmidt law
(Eq.~\ref{Sigma_SFR}) to an EM using Eq.~\ref{EM1} and equate the result
with the EM given in Eq.~\ref{EM2b} (with the expression in
Eq.~\ref{F_HII_cstar_b} substituted for $F_{\rm H~II}({\rm C~II}^*)$),
we find that
\begin{equation}
{Y_{n({\rm C~II}^*)}({\rm C~II})\over 1+0.057T_4^{-0.333}Y_{n({\rm
C~II}^*)}({\rm C~II})}\:=\:0.7^{+1.0}_{-0.4}\:T_4^{0.333}\:\Big[1-F_{\rm
H~II}({\rm S~II})\:\Big]
\end{equation}
(once again we have consolidated the slightly different kinds of
temperature terms $T_{4,n(e)^2}$ and $T_{4,n({\rm C~II}^*)}$ into a
single term $T_4$).  The uncertainty for the numerical coefficient on
the right-hand side of the equation considers the uncertainty in the
value of $\Sigma_{\rm SFR}$ combined with the uncertainties in the
column densities of H~I, S~II and \cstar\ that also appear in this
result.  Clearly, with $T_4=1$ we are unable to find an acceptable
result with the preferred value of the coefficient, since we require
that $Y_{n({\rm C~II}^*)}({\rm C~II})\geq 1.0$ and $F_{\rm H~II}({\rm
S~II})\geq 0$.  If this coefficient is at its (1$\sigma$) upper limit of
1.7 and $Y_{n({\rm C~II}^*)}({\rm C~II})=1.0$, then $F_{\rm H~II}({\rm
S~II})$ could equal 0.44.  However,  $Y_{n({\rm C~II}^*)}({\rm C~II})$ is
likely to be somewhat greater than 1.0, but by an amount that is
uncertain, so $F_{\rm H~II}({\rm S~II})$ is probably lower than 0.44.

To summarize, our consideration of the Schmidt law in conjunction with
our measurement of $N$(H~I) leads to an implied SFR which, when combined
with our interpretation on the origin of \cstar, places an upper limit
of about 44\% on the relative amount of S~II that could arise from H~II
regions.

While one might argue that in principle $\Sigma_{\rm SFR}$ could deviate above
the $+1 \sigma$ limit that was used to derive our upper limit for $F_{\rm H
  II}$(S II) if the conversion rate of gas into stars were significantly
higher than implied by the scaling of eq.(2), this possibility seems unlikely
for a galaxy with a surface brightness as low as that of SBS~1543+593.
The argument presented here, along with the one given in
\S\ref{ion_corr}, means that [S/H] for the gas we detected in
SBS~1543+593 is probably very close to the more naive determination
based on $\log[N({\rm S~II})/N({\rm H~I})]-\log$(S/H)$_\odot = -0.41$, but
it might be as low as $-0.66$.

\section{Physical Conditions of the Gas}\label{sect_physical}

If we knew the column density of C~II in the H~II region, we could
determine a representative electron density weighted by the density of
C~II inside it, i.e., $n(e)_{n({\rm C~II})}$.  This would be done
through the use of Eq.~\ref{ne1} (but with a replacement of local
densities $n$ with column densities $N$), because we know that most of
the \cstar\ arises from collisions with electrons within fully (or
mostly) ionized material.  Unfortunately, the information about $F_{\rm
H~II}({\rm S~II})$ cannot be applied to C, because the low value of
$F_{\rm H~II}({\rm S~II})$ could have arisen from the fact that $y({\rm
S~II})$ is very much greater than 1 over most parts of the region. 
Indeed, the models of \citet{stasinska90} indicate that S reverts to
multiply ionized stages more strongly than C, and this could in
principle help to explain the small value of $F_{\rm H~II}({\rm S~II})$.

For the neutral material, the situation is different.  Since it is
virtually certain that most of the C and S is singly ionized, we can
make the reasonable assumption that [C/S]~=~0 and state that inside the
H~I region
\begin{equation}
N({\rm C~II})\:=\:\left({\rm C\over S}\right)_\odot\:\Big[1-F_{\rm
H~II}({\rm S~II})\Big]\:N({\rm S~II})
\end{equation}
or
\begin{equation}
\log N({\rm C~II})\:=\:16.39~\:{\rm if~}F_{\rm H~II}({\rm S~II})\approx 0
\end{equation}
In \S\ref{sect_c1}, we reported a limit $\log N({\rm C~I})< 13.56$.  We
can use this limit together with our inferred value for $\log N({\rm
C~II})$ to place limits on $n(e)$ and $n({\rm H})$ in cool H~I regions
if we use the equilibrium equation for the balance of photoionizations,
at a rate $\Gamma_{\rm C}$, against recombinations with free electrons
(with a
rate constant $\alpha_e$) and negatively charged dust grains (with a
rate constant $\alpha_g$),
\begin{equation}\label{eqn_equilib}
[\alpha_en(e)+10^{[{\rm M/H}]}\alpha_gn({\rm H})]\:n({\rm
C~II})=\Gamma_{\rm C}n({\rm
C~I})\, .
\end{equation}
Formulae for the rate constants $\alpha_e$ and $\alpha_g$ are defined by
\citet{aldrovandi73} and \citet{weingartner01a},
respectively.  As in \S\ref{sect_FHII_cstar} above, when we evaluate
this equilibrium we assume that the
factor $10^{[{\rm M/H}]}$ reflects a reasonable estimate for the grain
density relative to that in our Galaxy.  The representative ionization
rate for neutral carbon
$\Gamma_{\rm C}$ should scale in proportion to $\Sigma_{\rm SFR}$ since
the atoms
are ionized by photons with energies greater than 11.26~eV.  Thus, once
again, we can scale a physical process using the Milky Way as an
example,
\begin{equation}
\Gamma_{\rm C}({\rm SBS~1543+593})=\Gamma_{\rm C}({\rm MW})\Sigma_{\rm
SFR}({\rm SBS~1543+593})/\Sigma_{\rm SFR}{\rm (MW)}\, ,
\end{equation}
where $\Gamma_{\rm C}{\rm (MW)}=2.24\times 10^{-10}{\rm s}^{-1}$
\citep{ebj_shaya79}.

Figure~\ref{fig_cicii} shows the upper bounds for the various
combinations
of $n(e)$ and $n({\rm H})$ that satisfy Eq.~\ref{eqn_equilib} and the
condition
that the observed $\log N({\rm C~I})-\log N({\rm C~II})<-2.83$,
representing the case $F_{\rm H~II}({\rm C~II})=0$, or the alternative
that $F_{\rm H~II}({\rm C~II})$ could be as low as our limit 0.44 for
$F_{\rm H~II}({\rm S~II})$.  In
effect, these two cases recognize that either a modest fraction or
virtually all of the observed C~II could arise from any H~I
region that could conceivably hold some C~I.

\begin{figure}[t]
\psfig{figure=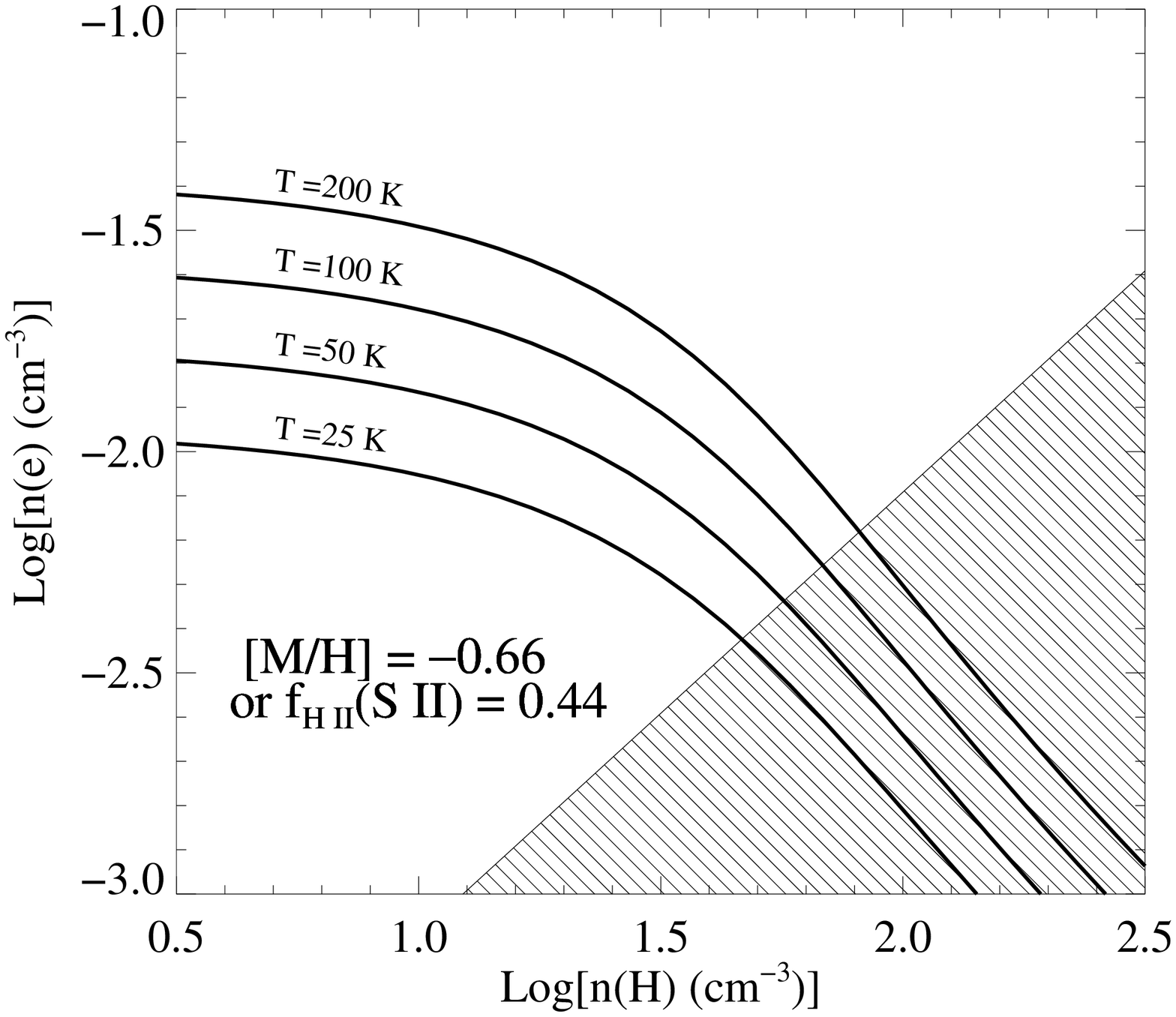,height=7cm} 
\vspace*{-7cm}\hspace*{8cm}\psfig{figure=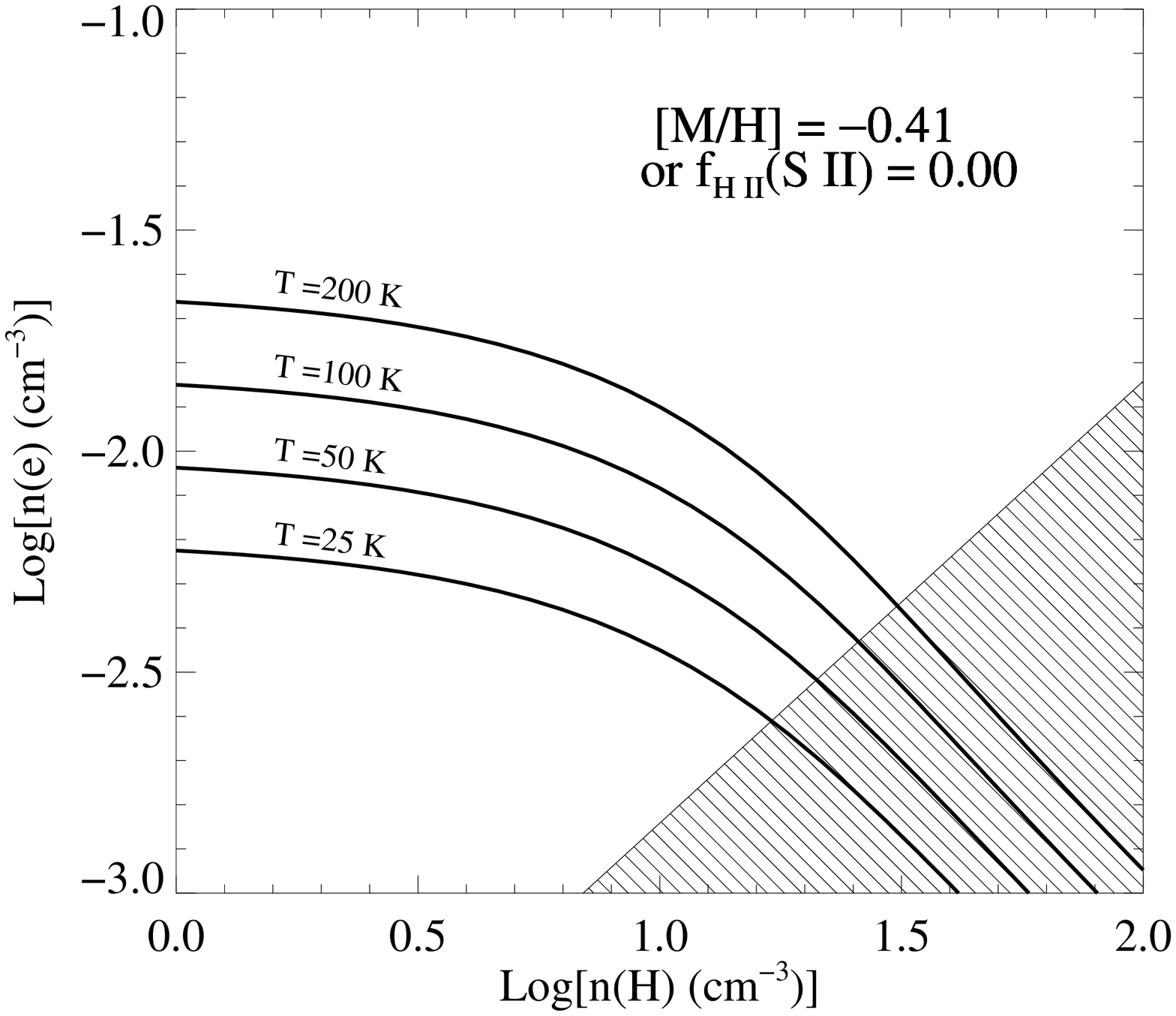,height=7cm}
\caption{\small 
Upper bounds that satisfy Eq.~\protect\ref{eqn_equilib} and our
observation that $\log N({\rm C~I})-\log N({\rm C~II})<-2.83$, i.e.,
combinations of $n(e)$ and $n({\rm H})$ below the curves for four
representative temperatures are the only ones allowed by our
observations.  The left-hand panel depicts the condition that only 0.56
times our assumed column density of C~II [derived from $N$(S~II) and our
assumption that [C/S]~=~0] is in the H~I region, while the right-hand
panel shows the conditions when virtually all of the C~II is in the H~I
region.  The shaded portion in the lower right portion of each panel
indicates a disallowed region where $n(e)$ is less than the product of
$n({\rm H})$ and the sum of the relative abundances of all elements that
can be photoionized in an H~I region. \label{fig_cicii}}
\end{figure}

\section{Summary \& Discussion} 

The primary goal of our STIS program was to obtain the abundance of sulfur and
nickel in the interstellar medium of the galaxy SBS~1543+593, by analyzing
ultraviolet absorption lines produced by the galaxy in the spectrum of the
background QSO HS~1543+5921.  With the good quality data obtained, and a
careful analysis of the absorption lines detected (including an optimal
extraction technique implemented to improve the standard pipeline extraction),
we find that the abundance of sulfur is approximately two-fifths solar,
$[\rm{S/H}] = -0.41 \pm 0.06$ (Table~\ref{tab_summ}).  This value is in good
agreement with the oxygen abundance, $[\rm{O/H}] = -0.49 \pm 0.2$ deduced by
\citet{regina04} and $[\rm{S/H}] \simeq\: 0.3 \pm 0.3$ measured by
\cite{regina05} from an analysis of nebular emission lines in the optical
spectrum of the brightest \ion{H}{2} region in SBS~1543+593.  The most
straightforward conclusion is that the two methods give a concordant picture
of the degree of metal enrichment in this galaxy, at least as far as the
$\alpha$-capture elements---which are mostly the products of Type~II
supernovae---are concerned.

\begin{deluxetable}{lccl}
\tabletypesize{\small}
\tablewidth{0pt}
\tablecaption{Summary of Abundances in SBS~1543+593 \label{tab_summ}}
\tablecolumns{4}
\tablehead{
\colhead{X}          &
\colhead{$\log $(X/H)$_{\odot}$\tablenotemark{a}} &
\colhead{} & 
\colhead{[X/H]}
}
\startdata
S  & $-4.81\pm0.04$ && \phantom{$<\,$} $-0.41\pm0.06$ \\
O  & $-3.31\pm0.05$ && $>\,-0.9$ \\
Si & $-4.46\pm0.02$ && $>\,-0.8$ \\
Ni & $-5.78\pm0.03$ && $<\,-0.81$ 
\enddata
\tablenotetext{a}{Reference solar abundances taken from the collation by \citet{lodder03}.}
\end{deluxetable}

As is often the case, there are caveats to this conclusion.  Emission and
absorption line measurements do not refer to exactly the same location within
the galaxy, but the two sight lines are separated by only 16.6 arcsec, or
3.3~\h\ at the redshift of the galaxy, and low luminosity galaxies like
SBS~1543+593 are found to exhibit relatively flat radial abundance gradients
\citep{deblok96}.  
In deriving the sulfur abundance in
the DLA, we have assumed that $N$(\ion{S}{2})/$N$(\ion{H}{1}) = (S/H). This
assumption would be incorrect if sulfur were depleted onto grains---a
possibility which we consider far-fetched given the composition of local
interstellar dust---or if some of the \ion{S}{2} absorption arises in ionized
gas which does not contribute to the damped \lya\ line.
If we have
under-estimated the total amount of hydrogen along the sight line, then the
measured abundance is only an upper limit.

Testing this last assumption has been a principle aim of this paper.
Since O~I tracks H~I so
closely in interstellar gas, an accurate measure of [O~I/S~II] would
enable us to measure the fraction of S~II arising in the observed
neutral hydrogen (assuming [O/S]$\:=\:0$).  Unfortunately, we have only a
lower limit to $N$(O~I) from the saturated O~I~$\lambda 1302$ line
(\S\ref{sect_o1}), and can show only that [O~I/S~II]$\:>\:-0.49$.
Hence we can only be certain that $>\:32$\% 
of the observed S~II line arises in neutral gas.  Even the lower limit
to $N$(O~I) is subject to uncertainty: extra components of high
velocity O~I (relative to the bulk of the absorption) could arrange themselves
in such as way that we adopt too narrow a Doppler parameter in calculating the
lower limit, leading us to over-estimate the O~I column density.

We can perhaps look to our own Galaxy to understand how the ionization stages
of sulfur track neutral hydrogen.  In the Milky Way, (S~II/H~I) ratios through
the disk are exactly [S/H]$_{\odot}$ \citep{lehner04} at the same H~I column
densities measured towards SBS~1543+593, suggesting that all the S~II arises
in Milky Way H~I gas (assuming that most of the ISM does not have super-solar
sulfur metallicities).  Although conditions in the ISM of SBS~1543+593 may be
different from that present in the Milky Way, it seems somewhat reassuring
that at least in one high-$N$(H~I) environment, most of the S~II comes from
H~I regions. 

Nevertheless, a more quantitative estimate of any required ionization
correction would be preferable.  Hence we have used the CLOUDY photoionization
code to estimate how much S~II arises in neutral or ionized regions. Not
surprisingly, we find that the answer depends on the shape of the UV
background field adopted for the calculation.  If a field similar in shape to
that of the Milky Way is used, then no ionization corrections are required.  A
much softer background though, or many embedded H~II regions, could leave some
of the S~II arising in H~II gas.  Given the distribution of candidate H~II
regions in the galaxy identified by \citet{regina04}, a Milky Way-like disk
model seems plausible, but we acknowledge that without a real measure of the
UV field near the center of the galaxy, CLOUDY models may only be a simplistic
approximation of the conditions in the interstellar gas.

Given all these difficulties, we have developed a different approach to
deriving an ionization correction, based on comparing two different methods
for deriving the emission measure of the ionized gas (\S\ref{sect_alternate}).
We have used the measured $N$(H~I) to estimate the star formation per unit
area in SBS~1543+593, and from there--- assuming none of the Lyman Limit photons
produced by the stars escape (or are absorbed by dust) --- we derive the emission
measure. We have also detected \cstar\ absorption in the ISM of the galaxy, which
enables us to calculate the same emission
measure. Combining these two results yields a relationship between the amount
of sulfur arising in ionized gas and the amount of carbon which exists as
C~II. Using plausible values for the latter value leads us to believe that no
more that 44\% of the S~II is likely to arise in ionized gas. At most,
therefore, the sulfur abundance might be as low as $-0.66$, only 0.25 dex
smaller than if we had assumed all the S~II came from neutral gas.

Since we do not detect the \ion{Ni}{2}~$\lambda 1317$ line, we can only deduce
an upper limit to the abundance of nickel, $[\rm{Ni/H}] < -0.81$ or ${\rm
  (Ni/H)} < 1/7$ of solar.  This limit implies that nickel is less abundant
than sulfur by a factor of 3 or more.  It seems likely that nickel is
underabundant due to depletion by dust, as seen in the local ISM, where it is
$\sim 10-300$ times below its solar value \citep[e.g.][]{ARAA_ghrs,
  ebj_carn04}. On the other hand, since nickel is an Fe-peak element
synthesized mostly by Type~Ia supernovae, it may be intrinsically less
abundant than sulfur (and oxygen). We might expect there to be little
over-abundance of alpha elements relative to iron-peak elements, since the
relatively high metallicity of sulfur should imply that sufficient time has
passed for Type Ia supernovae to build up the iron-peak abundances.  Of
course, both factors may contribute to the underabundance of nickel; in order
to assess their relative importance we need to measure an undepleted Fe-peak
element like zinc.

Assuming that our measured sulfur abundance is correct, it is reassuring to
confirm that emission- and absorption-based abundance measurements do give
consistent results, as we would expect.  Of course, the case reported here
refers to only one (late-type dwarf) galaxy but, as emphasized above, the test
we have carried out is the `cleanest' so far, free from the complications
introduced by: (i) differences in dust depletion and nucleosynthetic origin of
the elements being compared, (ii) absorption line saturation, and (iii) the
geometry of composite sight lines through the absorbing medium. Our result,
therefore, has implications for the interpretation of similar comparisons
which have been reported recently, as we now briefly discuss.

There seems to be no need for an envelope of unprocessed, neutral gas
surrounding this dwarf galaxy to contribute to the observed absorption, in
contrast with the suggestion put forward by \citet{cannon05} in their
interpretation of FUSE absorption line spectroscopy of a number of
star-bursting dwarf galaxies.  Possibly, this difference is related to the
evolutionary status of the galaxies involved---SBS~1543+593 is more metal-rich
than some, though not all, of the galaxies considered by \citet{cannon05} and
is \emph{not} undergoing a current burst of star formation.  One could think
of plausible scenarios where the existence of a metal-poor halo around
galaxies may be related to both of these factors.  Of course, measuring
abundances in dwarf galaxies with FUSE has its own set of unique
problems.  For one thing, absorption lines are seen against the background
light of the central starburst which is an extended, rather than point-like,
source.  In addition, the relatively small telescope aperture of FUSE
can only record such spectra at modest signal-to-noise ratios (S/N); in these
conditions the absorption lines studied tend to be strong and saturated.  Had
our metallicity measurements in SBS~1543+593 been based on \ion{O}{1}~$\lambda
1302$ and other similarly saturated absorption lines, we may well have
(erroneously) concluded that there is an order of magnitude difference between
\ion{H}{1} and \ion{H}{2} region abundances in this galaxy too
(\S\ref{sect_o1}).

\begin{figure}[t]
\vspace*{0cm}\centerline{\psfig
{figure=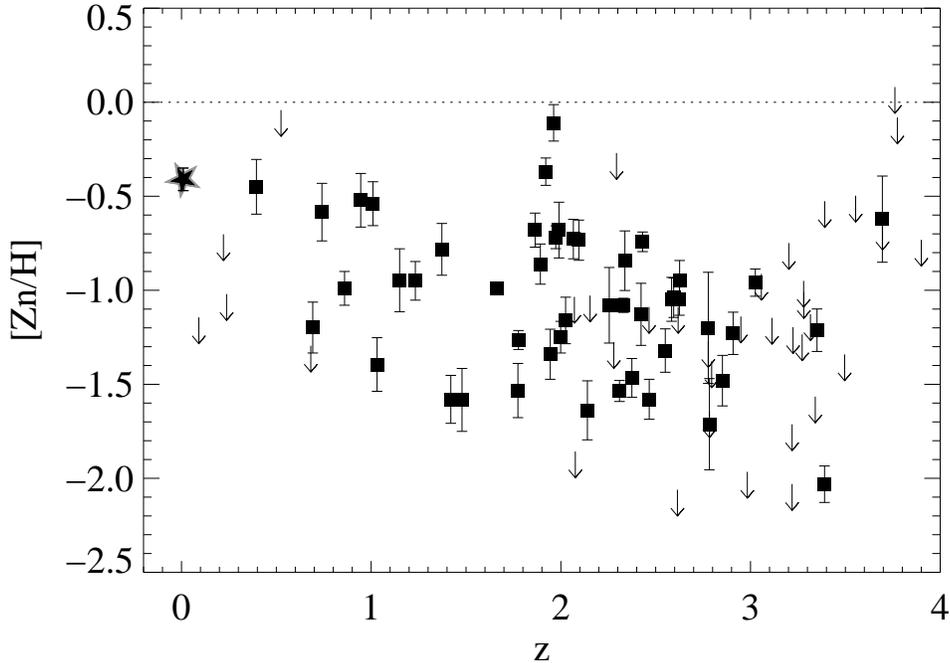,height=10cm}}
\caption{\small The current sample of zinc abundance measurements 
 in DLAs, as a function of redshift $z$, from the compilation by \citet{kulkarni05}. 
 The star close to $z=0$ represents the sulfur abundance we have measured
 in SBS~1543+593. Any offset between metallicities measured from zinc and sulfur is
 likely to be less than 0.2\,dex (see Figure~9 of \citealt{nissen04}).
 \label{fig_dlas}}
 \end{figure}

As for the population of DLAs as a whole,  it can be seen from Figure~\ref{fig_dlas}
that SBS~1543+593 is among the least metal-poor DLAs
at $z < 1$ (and indeed, at any redshift). This may well be 
the result of the unusually low impact parameter of this
QSO-galaxy pairing, as the sight line to HS~1543+5921
intersects the galaxy at $\leq\: 0.5$~\h\ from its center.
In other low-$z$ DLAs, radial metallicity gradients have been
proposed to explain the finding that [Fe/H]$_{\rm DLA}$ is
lower than [O/H]$_{\rm H~II}$ \citep{chen05}.
We would have found a similar result 
ourselves, had we based our abundance comparison on
nickel (\S\ref{sect_NiII}) which, like iron, can be depleted from 
the gas-phase of the ISM by significant, and variable, amounts.
Furthermore, the systematic offsets which have recently
been identified in the derivation of [O/H]
from the ratios of strong nebular emission lines 
\citep{garnett04_m51, garnett04_te, bresolin04}
seem to be less important when the oxygen abundance
is sub-solar, as in SBS~1543+5921,
than solar or super-solar, as is the case for 
the galaxies considered by \citet{chen05}.
These offsets would work in the sense
of increasing the difference between \ion{H}{2} and 
\ion{H}{1} region abundances.
Thus, in order to assess the degree to which
abundance gradients are responsible for the generally low
metallicities of DLAs, we need: (i) better determinations
of the gradients in local galaxies and (ii) more instances like
the one reported here where emission- and absorption-deduced
abundances can be compared without the uncertainties
associated with dust depletion and strong emission line 
calibrators. 

The cooling rates $l_c$ of the DLAs derived from \cstar\ absorption lines have
been investigated in detail by \citet{wolfe03a,wolfe03b,wolfe04}.  The median
value of $\log( l_c)$ in the DLA population is $\simeq -27.00$ (where $l_c$ is
in units of ergs s$^{-1}$ H~atom$^{-1}$). Calculating $l_c$ for SBS~1543+593
relies on knowing the fraction of \cstar\ in H~II gas, i.e., $F_{\rm H~II}$(\cstar )
in eqn~(\ref{l_C}). We can use plausible values of $Y_{n({\rm C~II}^*)}({\rm
  C~II})$ to estimate $F_{\rm H~II}$(\cstar ) via eqn~\ref{F_HII_cstar_b}: for
the range $Y_{n({\rm C~II}^*)}({\rm C~II})\:=\:$[1,22] (see
\S\ref{sect_cstarHII}), we find $\log (l_c)\:=\:[-26.9,-25.9]$. 
However, if the star formation rate in SBS~1543+593 is close to the value
suggested in \S\ref{sect_SFR_HI}, $\simeq 1-2\times
10^{-3}$~M$_\odot$~yr$^{-1}$~kpc$^{-2}$, then: $Y_{n({\rm C~II}^*)}({\rm
  C~II})$ must be between $\approx 1-2$; the fraction of S~II which resides in
ionized gas must be $<<$~44\%; the fraction of \cstar\ arising in ionized gas
would be $>90$~\%; and the cooling rate would be close to
the median value of  $\log(l_c)\:=\:-27.0$ for the DLAs.

Whether such a comparison is entirely valid is, however, unclear.  For the
DLAs, \cstar\ absorption is thought to come primarily from the cold {\it
  neutral} medium, i.e. with $F_{\rm H~II}$(\cstar )$\:\simeq\:0$
\citep{wolfe04}. Unfortunately, a more detailed comparison between the methods
used to measure $l_c$ in this paper and those used for the DLA population is
beyond the scope of this paper.

Our observations of SBS~1543+593
provide a strong incentive for finding more
examples of close QSO-galaxy pairs at low redshifts, where the galaxy
producing the DLA in the QSO spectrum can be imaged and its
properties measured. While on-going large scale surveys, such
as the \emph{Sloan Digital Sky Survey}, have dramatically 
increased the numbers of such pairs, the current lack of a
sensitive, high resolution, near-UV spectrograph in space
will hamper progress in this area until a worthy successor of STIS 
is flown.

\acknowledgments 

We thank Bruce Draine for an important reading of a late draft of this
paper. Support for this project was provided by NASA through grant
HST-GO-09784 from the Space Telescope Science Institute, which is operated by
the Association of Universities for Research in Astronomy, Inc., under NASA
contract NAS5-26555. T.M.T. appreciates support from NASA grant NNG 04GG73G.

\bibliographystyle{apj}
\bibliography{apj-jour,bib2}

\end{document}